\input harvmac

%\draftmode

\input amssym

%\DijkgraafPZ
\lref\DijkgraafPZ{
  R.~Dijkgraaf and E.~Witten,
  ``Topological Gauge Theories And Group Cohomology,''
  Commun.\ Math.\ Phys.\  {\bf 129}, 393 (1990).
  %%CITATION = CMPHA,129,393;%%
}

%\BelovZE
\lref\BelovZE{
  D.~Belov and G.~W.~Moore,
  ``Classification of abelian spin Chern-Simons theories,''
  arXiv:hep-th/0505235.
  %%CITATION = HEP-TH/0505235;%%
}

%\KapustinHK
\lref\KapustinHK{
  A.~Kapustin and N.~Saulina,
  ``Topological boundary conditions in abelian Chern-Simons theory,''
  arXiv:1008.0654 [hep-th].
  %%CITATION = ARXIV:1008.0654;%%
}

%\WittenCD
\lref\WittenCD{
  E.~Witten,
  ``D-branes and K-theory,''
  JHEP {\bf 9812}, 019 (1998)
  [arXiv:hep-th/9810188].
  %%CITATION = JHEPA,9812,019;%%
}
%\SenRG
\lref\SenRG{
  A.~Sen,
  ``Stable non-BPS states in string theory,''
  JHEP {\bf 9806}, 007 (1998)
  [arXiv:hep-th/9803194].
  %%CITATION = JHEPA,9806,007;%%
}

%\SenII
\lref\SenII{
  A.~Sen,
  ``Stable non-BPS bound states of BPS D-branes,''
  JHEP {\bf 9808}, 010 (1998)
  [arXiv:hep-th/9805019].
  %%CITATION = JHEPA,9808,010;%%
}

%\MinasianMM
\lref\MinasianMM{
  R.~Minasian and G.~W.~Moore,
  ``K-theory and Ramond-Ramond charge,''
  JHEP {\bf 9711}, 002 (1997)
  [arXiv:hep-th/9710230].
  %%CITATION = JHEPA,9711,002;%%
}

%\SenSM
\lref\SenSM{
  A.~Sen,
  ``Tachyon condensation on the brane antibrane system,''
  JHEP {\bf 9808}, 012 (1998)
  [arXiv:hep-th/9805170].
  %%CITATION = JHEPA,9808,012;%%
}

%\BanksYZ
\lref\BanksYZ{
  T.~Banks and L.~J.~Dixon,
  ``Constraints on String Vacua with Space-Time Supersymmetry,''
  Nucl.\ Phys.\  B {\bf 307}, 93 (1988).
  %%CITATION = NUPHA,B307,93;%%
}

%\PolchinskiBQ
\lref\PolchinskiBQ{
  J.~Polchinski,
  ``Monopoles, duality, and string theory,''
  Int.\ J.\ Mod.\ Phys.\  A {\bf 19S1}, 145 (2004)
  [arXiv:hep-th/0304042].
  %%CITATION = IMPAE,A19S1,145;%%
}
%\DouglasUP
\lref\DouglasUP{
  M.~R.~Douglas, I.~R.~Klebanov, D.~Kutasov, J.~M.~Maldacena, E.~J.~Martinec and N.~Seiberg,
  ``A new hat for the c = 1 matrix model,''
  arXiv:hep-th/0307195.
  %%CITATION = HEP-TH/0307195;%%
}
%\KlebanovWG
\lref\KlebanovWG{
  I.~R.~Klebanov, J.~M.~Maldacena and N.~Seiberg,
  ``Unitary and complex matrix models as 1-d type 0 strings,''
  Commun.\ Math.\ Phys.\  {\bf 252}, 275 (2004)
  [arXiv:hep-th/0309168].
  %%CITATION = CMPHA,252,275;%%
}
%\MaldacenaHE
\lref\MaldacenaHE{
  J.~M.~Maldacena and N.~Seiberg,
  ``Flux-vacua in two dimensional string theory,''
  JHEP {\bf 0509}, 077 (2005)
  [arXiv:hep-th/0506141].
  %%CITATION = JHEPA,0509,077;%%
}
%\SeibergEI
\lref\SeibergEI{
  N.~Seiberg and D.~Shih,
  ``Flux vacua and branes of the minimal superstring,''
  JHEP {\bf 0501}, 055 (2005)
  [arXiv:hep-th/0412315].
  %%CITATION = JHEPA,0501,055;%%
}
%\SeibergNM
\lref\SeibergNM{
  N.~Seiberg and D.~Shih,
  ``Branes, rings and matrix models in minimal (super)string theory,''
  JHEP {\bf 0402}, 021 (2004)
  [arXiv:hep-th/0312170].
  %%CITATION = JHEPA,0402,021;%%
}

%\PolchinskiBG
\lref\PolchinskiBG{
  J.~Polchinski,
  ``Open heterotic strings,''
  JHEP {\bf 0609}, 082 (2006)
  [arXiv:hep-th/0510033].
  %%CITATION = JHEPA,0609,082;%%
}

\lref\nima{{\bf What is this reference?}}

\lref\bkmpeskin{BKMPeskin  {\bf What is this reference?}}

\lref\cepsguw{cepsguw  {\bf What is this reference?}}

%\KraussZC
\lref\KraussZC{
  L.~M.~Krauss and F.~Wilczek,
  ``Discrete Gauge Symmetry in Continuum Theories,''
  Phys.\ Rev.\ Lett.\  {\bf 62}, 1221 (1989).
  %%CITATION = PRLTA,62,1221;%%
}
%\AlfordSJ
\lref\AlfordSJ{
  M.~G.~Alford and F.~Wilczek,
  ``Aharonov-Bohm Interaction of Cosmic Strings with Matter,''
  Phys.\ Rev.\ Lett.\  {\bf 62}, 1071 (1989).
  %%CITATION = PRLTA,62,1071;%%
}

%\PreskillBM
\lref\PreskillBM{
  J.~Preskill and L.~M.~Krauss,
  ``Local Discrete Symmetry And Quantum Mechanical Hair,''
  Nucl.\ Phys.\  B {\bf 341}, 50 (1990).
  %%CITATION = NUPHA,B341,50;%%
}

%\HorowitzNG
\lref\HorowitzNG{
  G.~T.~Horowitz,
  ``Exactly Soluble Diffeomorphism Invariant Theories,''
  Commun.\ Math.\ Phys.\  {\bf 125}, 417 (1989).
  %%CITATION = CMPHA,125,417;%%
}

%\MaldacenaSS
\lref\MaldacenaSS{
  J.~M.~Maldacena, G.~W.~Moore and N.~Seiberg,
  ``D-brane charges in five-brane backgrounds,''
  JHEP {\bf 0110}, 005 (2001)
  [arXiv:hep-th/0108152].
  %%CITATION = JHEPA,0110,005;%%
}

%\CaldararuTC
\lref\CaldararuTC{
  A.~Caldararu, J.~Distler, S.~Hellerman, T.~Pantev and E.~Sharpe,
  ``Non-birational twisted derived equivalences in abelian GLSMs,''
  arXiv:0709.3855 [hep-th].
  %%CITATION = ARXIV:0709.3855;%%
}

%\GarfinkleQJ
\lref\GarfinkleQJ{
  D.~Garfinkle, G.~T.~Horowitz and A.~Strominger,
  ``Charged black holes in string theory,''
  Phys.\ Rev.\  D {\bf 43}, 3140 (1991)
  [Erratum-ibid.\  D {\bf 45}, 3888 (1992)]
  [Phys.\ Rev.\  D {\bf 45}, 3888 (1992)].
  %%CITATION = PHRVA,D45,3888;%%
}

%\HorowitzCD
\lref\HorowitzCD{
  G.~T.~Horowitz and A.~Strominger,
  %``Black strings and P-branes,''
  Nucl.\ Phys.\  B {\bf 360}, 197 (1991).
  %%CITATION = NUPHA,B360,197;%%
}

%\PantevRH
\lref\PantevRH{
  T.~Pantev and E.~Sharpe,
  ``Notes on gauging noneffective group actions,''
  arXiv:hep-th/0502027.
  %%CITATION = HEP-TH/0502027;%%
}

%\PantevZS
\lref\PantevZS{
  T.~Pantev and E.~Sharpe,
  ``GLSM's for gerbes (and other toric stacks),''
  Adv.\ Theor.\ Math.\ Phys.\  {\bf 10}, 77 (2006)
  [arXiv:hep-th/0502053].
  %%CITATION = 00203,10,77;%%
}

%\SeibergQD
\lref\SeibergQD{
  N.~Seiberg,
  ``Modifying the Sum Over Topological Sectors and Constraints on
  Supergravity,''
  JHEP {\bf 1007}, 070 (2010)
  [arXiv:1005.0002 [hep-th]].
  %%CITATION = JHEPA,1007,070;%%
}
 % EXTERNAL FILES

%\SohniusTP
\lref\SohniusTP{
  M.~F.~Sohnius and P.~C.~West,
  ``An Alternative Minimal Off-Shell Version Of N=1 Supergravity,''
  Phys.\ Lett.\  B {\bf 105}, 353 (1981).
  %%CITATION = PHLTA,B105,353;%%
}

%\GatesNR
\lref\GatesNR{
  S.~J.~Gates, M.~T.~Grisaru, M.~Rocek and W.~Siegel,
  ``Superspace, or one thousand and one lessons in supersymmetry,''
  Front.\ Phys.\  {\bf 58}, 1 (1983)
  [arXiv:hep-th/0108200].
  %%CITATION = FRPHA,58,1;%%
}

%\KomargodskiPC
\lref\KomargodskiPC{
  Z.~Komargodski and N.~Seiberg,
  ``Comments on the Fayet-Iliopoulos Term in Field Theory and Supergravity,''
  JHEP {\bf 0906}, 007 (2009)
  [arXiv:0904.1159 [hep-th]].
  %%CITATION = JHEPA,0906,007;%%
}

%\ArkaniHamedRS
\lref\ArkaniHamedRS{
  N.~Arkani-Hamed, S.~Dimopoulos and G.~R.~Dvali,
  ``The hierarchy problem and new dimensions at a millimeter,''
  Phys.\ Lett.\  B {\bf 429}, 263 (1998)
  [arXiv:hep-ph/9803315].
  %%CITATION = PHLTA,B429,263;%%
}

%\LambertDX
\lref\LambertDX{
  N.~D.~Lambert and G.~W.~Moore,
  ``Distinguishing off-shell supergravities with on-shell physics,''
  Phys.\ Rev.\  D {\bf 72}, 085018 (2005)

  [arXiv:hep-th/0507018].
  %%CITATION = PHRVA,D72,085018;%%
}

%\AkulovCK
\lref\AkulovCK{
  V.~P.~Akulov, D.~V.~Volkov and V.~A.~Soroka,
  ``On The General Covariant Theory Of Calibrating Poles In Superspace,''
  Theor.\ Math.\ Phys.\  {\bf 31}, 285 (1977)
  [Teor.\ Mat.\ Fiz.\  {\bf 31}, 12 (1977)].
  %%CITATION = TMFZA,31,12;%%
}

%\KalloshVE
\lref\KalloshVE{
  R.~Kallosh, L.~Kofman, A.~D.~Linde and A.~Van Proeyen,
  ``Superconformal symmetry, supergravity and cosmology,''
  Class.\ Quant.\ Grav.\  {\bf 17}, 4269 (2000)
  [Erratum-ibid.\  {\bf 21}, 5017 (2004)]
  [arXiv:hep-th/0006179].
  %%CITATION = CQGRD,17,4269;%%
}

%\DvaliZH
\lref\DvaliZH{
  G.~Dvali, R.~Kallosh and A.~Van Proeyen,
  ``D-term strings,''
  JHEP {\bf 0401}, 035 (2004)
  [arXiv:hep-th/0312005].
  %%CITATION = JHEPA,0401,035;%%
}

%\ColemanUZ
\lref\ColemanUZ{
  S.~R.~Coleman,
  ``More About The Massive Schwinger Model,''
  Annals Phys.\  {\bf 101}, 239 (1976).
  %%CITATION = APNYA,101,239;%%
}

%\FischlerZK
\lref\FischlerZK{
  W.~Fischler, H.~P.~Nilles, J.~Polchinski, S.~Raby and L.~Susskind,
  ``Vanishing Renormalization Of The D Term In Supersymmetric U(1) Theories,''
  Phys.\ Rev.\ Lett.\  {\bf 47}, 757 (1981).
  %%CITATION = PRLTA,47,757;%%
}

%\WittenNF
\lref\WittenNF{
  E.~Witten,
  ``Dynamical Breaking Of Supersymmetry,''
  Nucl.\ Phys.\  B {\bf 188}, 513 (1981).
  %%CITATION = NUPHA,B188,513;%%
}

\lref\Wittenun{E.~Witten, unpublished}

%\Distlerun
\lref\Distlerun{
  J.~Distler and B.~Wecht,
  Unpublished, mentioned in {\sl http://golem.ph.utexas.edu/~distler
  /blog/archives/002180.html  }
  }
%\GreeneYA
\lref\GreeneYA{
  B.~R.~Greene, A.~D.~Shapere, C.~Vafa and S.~T.~Yau,
  ``Stringy Cosmic Strings And Noncompact Calabi-Yau Manifolds,''
  Nucl.\ Phys.\  B {\bf 337}, 1 (1990).
  %%CITATION = NUPHA,B337,1;%%
}
%\AshokGK
\lref\AshokGK{
  S.~Ashok and M.~R.~Douglas,
  ``Counting flux vacua,''
  JHEP {\bf 0401}, 060 (2004)
  [arXiv:hep-th/0307049].
  %%CITATION = JHEPA,0401,060;%%
}

%\DistlerZG
\lref\DistlerZG{
  J.~Distler and E.~Sharpe,
  ``Quantization of Fayet-Iliopoulos Parameters in Supergravity,''
  arXiv:1008.0419 [hep-th].
  %%CITATION = ARXIV:1008.0419;%%
}

%\ElvangJK
\lref\ElvangJK{
  H.~Elvang, D.~Z.~Freedman and B.~Kors,
  ``Anomaly cancellation in supergravity with Fayet-Iliopoulos couplings,''
  JHEP {\bf 0611}, 068 (2006)
  [arXiv:hep-th/0606012].
  %%CITATION = JHEPA,0611,068;%%
}

%\ShifmanZI
\lref\ShifmanZI{
  M.~A.~Shifman and A.~I.~Vainshtein,
  ``Solution of the Anomaly Puzzle in SUSY Gauge Theories and the Wilson
  Operator Expansion,''
  Nucl.\ Phys.\  B {\bf 277}, 456 (1986)
  [Sov.\ Phys.\ JETP {\bf 64}, 428 (1986\ ZETFA,91,723-744.1986)].
  %%CITATION = ZETFA,91,723;%%
}

%\MagroAJ
\lref\MagroAJ{
  M.~Magro, I.~Sachs and S.~Wolf,
  ``Superfield Noether procedure,''
  Annals Phys.\  {\bf 298}, 123 (2002)
  [arXiv:hep-th/0110131].
  %%CITATION = APNYA,298,123;%%
}

%\KuzenkoNI
\lref\KuzenkoNI{
  S.~M.~Kuzenko,
  ``Variant supercurrents and Noether procedure,''
  arXiv:1008.1877 [hep-th].
  %%CITATION = ARXIV:1008.1877;%%
}

%\KuzenkoYM
\lref\KuzenkoYM{
  S.~M.~Kuzenko,
  ``The Fayet-Iliopoulos term and nonlinear self-duality,''
  arXiv:0911.5190 [hep-th].
  %%CITATION = ARXIV:0911.5190;%%
}

%\KuzenkoAM
\lref\KuzenkoAM{
  S.~M.~Kuzenko,
  ``Variant supercurrent multiplets,''
  JHEP {\bf 1004}, 022 (2010)
  [arXiv:1002.4932 [hep-th]].
  %%CITATION = JHEPA,1004,022;%%
}

%\KuzenkoNI
\lref\KuzenkoNI{
  S.~M.~Kuzenko,
  ``Variant supercurrents and Noether procedure,''
  arXiv:1008.1877 [hep-th].
  %%CITATION = ARXIV:1008.1877;%%
}

%\LambertDX
\lref\LambertDX{
  N.~D.~Lambert and G.~W.~Moore,
  ``Distinguishing off-shell supergravities with on-shell physics,''
  Phys.\ Rev.\  D {\bf 72}, 085018 (2005)
  [arXiv:hep-th/0507018].
  %%CITATION = PHRVA,D72,085018;%%
}

%\KuzenkoAM
\lref\KuzenkoAM{
  S.~M.~Kuzenko,
  ``Variant supercurrent multiplets,''
  JHEP {\bf 1004}, 022 (2010)
  [arXiv:1002.4932 [hep-th]].
  %%CITATION = JHEPA,1004,022;%%
}

%\FreedBN
\lref\FreedBN{
  D.~S.~Freed and F.~Quinn,
  ``Chern-Simons theory with finite gauge group,''
  Commun.\ Math.\ Phys.\  {\bf 156}, 435 (1993)
  [arXiv:hep-th/9111004].
  %%CITATION = CMPHA,156,435;%%
}

%\MisnerMT
\lref\MisnerMT{
  C.~W.~Misner and J.~A.~Wheeler,
  ``Classical physics as geometry: Gravitation, electromagnetism, unquantized
  %charge, and mass as properties of curved empty space,''
  Annals Phys.\  {\bf 2}, 525 (1957).
  %%CITATION = APNYA,2,525;%%
}

%\BarbieriAC
\lref\BarbieriAC{
  R.~Barbieri, S.~Ferrara, D.~V.~Nanopoulos and K.~S.~Stelle,
  ``Supergravity, R Invariance And Spontaneous Supersymmetry Breaking,''
  Phys.\ Lett.\  B {\bf 113}, 219 (1982).
  %%CITATION = PHLTA,B113,219;%%
}

%\SiegelSV
\lref\SiegelSV{
  W.~Siegel,
  ``16/16 Supergravity,''
  Class.\ Quant.\ Grav.\  {\bf 3}, L47 (1986).
  %%CITATION = CQGRD,3,L47;%%
}

%\KuzenkoAM
\lref\KuzenkoAM{
  S.~M.~Kuzenko,
  ``Variant supercurrent multiplets,''
  arXiv:1002.4932 [hep-th].
  %%CITATION = ARXIV:1002.4932;%%
}

%\GirardiVQ
\lref\GirardiVQ{
  G.~Girardi, R.~Grimm, M.~Muller and J.~Wess,
  ``Antisymmetric Tensor Gauge Potential In Curved Superspace And A (16+16)
  Supergravity Multiplet,''
  Phys.\ Lett.\  B {\bf 147}, 81 (1984).
  %%CITATION = PHLTA,B147,81;%%
}

%\LangXK
\lref\LangXK{
  W.~Lang, J.~Louis and B.~A.~Ovrut,
  ``(16+16) Supergravity Coupled To Matter: The Low-Energy Limit Of The
  Superstring,''
  Phys.\ Lett.\  B {\bf 158}, 40 (1985).
  %%CITATION = PHLTA,B158,40;%%
}

%\CaldararuTC
\lref\CaldararuTC{
  A.~Caldararu, J.~Distler, S.~Hellerman, T.~Pantev and E.~Sharpe,
  ``Non-birational twisted derived equivalences in abelian GLSMs,''
  arXiv:0709.3855 [hep-th].
  %%CITATION = ARXIV:0709.3855;%%
}

%\SeibergVC
\lref\SeibergVC{
  N.~Seiberg,
  ``Naturalness Versus Supersymmetric Non-renormalization Theorems,''
  Phys.\ Lett.\  B {\bf 318}, 469 (1993)
  [arXiv:hep-ph/9309335].
  %%CITATION = PHLTA,B318,469;%%
}

%\DineXK
\lref\DineXK{
  M.~Dine, N.~Seiberg and E.~Witten,
  ``Fayet-Iliopoulos Terms in String Theory,''
  Nucl.\ Phys.\  B {\bf 289}, 589 (1987).
  %%CITATION = NUPHA,B289,589;%%
}

%\FreedmanUK
\lref\FreedmanUK{
  D.~Z.~Freedman,
  ``Supergravity With Axial Gauge Invariance,''
  Phys.\ Rev.\  D {\bf 15}, 1173 (1977).
  %%CITATION = PHRVA,D15,1173;%%
}

%\DasPU
\lref\DasPU{
  A.~Das, M.~Fischler and M.~Rocek,
  ``Superhiggs Effect In A New Class Of Scalar Models And A Model Of Super
  QED,''
  Phys.\ Rev.\  D {\bf 16}, 3427 (1977).
  %%CITATION = PHRVA,D16,3427;%%
}

     %\deWitWW
      \lref\deWitWW{
        B.~de Wit and P.~van Nieuwenhuizen,
       ``The Auxiliary Field Structure In Chirally Extended Supergravity,''
        Nucl.\ Phys.\  B {\bf 139}, 216 (1978).
        %%CITATION = NUPHA,B139,216;%%
      }

%\BinetruyHH
\lref\BinetruyHH{
  P.~Binetruy, G.~Dvali, R.~Kallosh and A.~Van Proeyen,
  ``Fayet-Iliopoulos terms in supergravity and cosmology,''
  Class.\ Quant.\ Grav.\  {\bf 21}, 3137 (2004)
  [arXiv:hep-th/0402046].
  %%CITATION = CQGRD,21,3137;%%
}

%\FerraraDH
\lref\FerraraDH{
  S.~Ferrara, L.~Girardello, T.~Kugo and A.~Van Proeyen,
  ``Relation Between Different Auxiliary Field Formulations Of N=1 Supergravity
  Coupled To Matter,''
  Nucl.\ Phys.\  B {\bf 223}, 191 (1983).
  %%CITATION = NUPHA,B223,191;%%
}

%\KuzenkoYM
\lref\KuzenkoYM{
  S.~M.~Kuzenko,
  ``The Fayet-Iliopoulos term and nonlinear self-duality,''
  arXiv:0911.5190 [hep-th].
  %%CITATION = ARXIV:0911.5190;%%
}

%\ClarkJX
\lref\ClarkJX{
  T.~E.~Clark, O.~Piguet and K.~Sibold,
  ``Supercurrents, Renormalization And Anomalies,''
  Nucl.\ Phys.\  B {\bf 143}, 445 (1978).
  %%CITATION = NUPHA,B143,445;%%
}

%\KomargodskiPC
\lref\KomargodskiPC{
  Z.~Komargodski and N.~Seiberg,
  ``Comments on the Fayet-Iliopoulos Term in Field Theory and Supergravity,''
  JHEP {\bf 0906}, 007 (2009)
  [arXiv:0904.1159 [hep-th]].
  %%CITATION = JHEPA,0906,007;%%
}

%\AffleckVC
\lref\AffleckVC{
  I.~Affleck, M.~Dine and N.~Seiberg,
  ``Dynamical Supersymmetry Breaking In Chiral Theories,''
  Phys.\ Lett.\ B {\bf 137}, 187 (1984).
  %%CITATION = PHLTA,B137,187;%%
}

%\GatesNR
\lref\GatesNR{
  S.~J.~Gates, M.~T.~Grisaru, M.~Rocek and W.~Siegel,
  ``Superspace, or one thousand and one lessons in supersymmetry,''
  Front.\ Phys.\  {\bf 58}, 1 (1983)
  [arXiv:hep-th/0108200].
  %%CITATION = FRPHA,58,1;%%
}

\lref\RocekPC{M.~Rocek, Private communication.}

%\SeibergBZ
\lref\SeibergBZ{
  N.~Seiberg,
  ``Exact results on the space of vacua of four-dimensional SUSY gauge
  theories,''
  Phys.\ Rev.\ D {\bf 49}, 6857 (1994)
  [arXiv:hep-th/9402044].
  %%CITATION = HEP-TH 9402044;%%
  }

%\KalloshVE
\lref\KalloshVE{
  R.~Kallosh, L.~Kofman, A.~D.~Linde and A.~Van Proeyen,
  ``Superconformal symmetry, supergravity and cosmology,''
  Class.\ Quant.\ Grav.\  {\bf 17}, 4269 (2000)
  [Erratum-ibid.\  {\bf 21}, 5017 (2004)]
  [arXiv:hep-th/0006179].
  %%CITATION = CQGRD,17,4269;%%
}

%\DvaliZH
\lref\DvaliZH{
  G.~Dvali, R.~Kallosh and A.~Van Proeyen,
  ``D-term strings,''
  JHEP {\bf 0401}, 035 (2004)
  [arXiv:hep-th/0312005].
  %%CITATION = JHEPA,0401,035;%%
}

%\FischlerZK
\lref\FischlerZK{
  W.~Fischler, H.~P.~Nilles, J.~Polchinski, S.~Raby and L.~Susskind,
  ``Vanishing Renormalization Of The D Term In Supersymmetric U(1) Theories,''
  Phys.\ Rev.\ Lett.\  {\bf 47}, 757 (1981).
  %%CITATION = PRLTA,47,757;%%
}

%\WittenNF
\lref\WittenNF{
  E.~Witten,
  ``Dynamical Breaking Of Supersymmetry,''
  Nucl.\ Phys.\  B {\bf 188}, 513 (1981).
  %%CITATION = NUPHA,B188,513;%%
}

%\NemeschanskyYX
\lref\NemeschanskyYX{
  D.~Nemeschansky and A.~Sen,
  ``Conformal Invariance of Supersymmetric Sigma Models on Calabi-Yau
  Manifolds,''
  Phys.\ Lett.\  B {\bf 178}, 365 (1986).
  %%CITATION = PHLTA,B178,365;%%
}

\lref\EW{E.~Witten, unpublished.}

\lref\KS{Z.~Komargodski and N.~Seiberg, to appear.}

%\ElvangJK
\lref\ElvangJK{
  H.~Elvang, D.~Z.~Freedman and B.~Kors,
  ``Anomaly cancellation in supergravity with Fayet-Iliopoulos couplings,''
  JHEP {\bf 0611}, 068 (2006)
  [arXiv:hep-th/0606012].
  %%CITATION = JHEPA,0611,068;%%
}

%\ShifmanZI
\lref\ShifmanZI{
  M.~A.~Shifman and A.~I.~Vainshtein,
  ``Solution of the Anomaly Puzzle in SUSY Gauge Theories and the Wilson
  Operator Expansion,''
  Nucl.\ Phys.\  B {\bf 277}, 456 (1986)
  [Sov.\ Phys.\ JETP {\bf 64}, 428 (1986\ ZETFA,91,723-744.1986)].
  %%CITATION = ZETFA,91,723;%%
}

%\BarbieriAC
\lref\BarbieriAC{
  R.~Barbieri, S.~Ferrara, D.~V.~Nanopoulos and K.~S.~Stelle,
  ``Supergravity, R Invariance And Spontaneous Supersymmetry Breaking,''
  Phys.\ Lett.\  B {\bf 113}, 219 (1982).
  %%CITATION = PHLTA,B113,219;%%
}

%\SeibergVC
\lref\SeibergVC{
  N.~Seiberg,
  ``Naturalness Versus Supersymmetric Non-renormalization Theorems,''
  Phys.\ Lett.\  B {\bf 318}, 469 (1993)
  [arXiv:hep-ph/9309335].
  %%CITATION = PHLTA,B318,469;%%
}

%\DineXK
\lref\DineXK{
  M.~Dine, N.~Seiberg and E.~Witten,
  ``Fayet-Iliopoulos Terms in String Theory,''
  Nucl.\ Phys.\  B {\bf 289}, 589 (1987).
  %%CITATION = NUPHA,B289,589;%%
}

%\FreedmanUK
\lref\FreedmanUK{
  D.~Z.~Freedman,
  ``Supergravity With Axial Gauge Invariance,''
  Phys.\ Rev.\  D {\bf 15}, 1173 (1977).
  %%CITATION = PHRVA,D15,1173;%%
}

%\DasPU
\lref\DasPU{
  A.~Das, M.~Fischler and M.~Rocek,
  ``Superhiggs Effect In A New Class Of Scalar Models And A Model Of Super
  QED,''
  Phys.\ Rev.\  D {\bf 16}, 3427 (1977).
  %%CITATION = PHRVA,D16,3427;%%
}

%\BinetruyHH
\lref\BinetruyHH{
  P.~Binetruy, G.~Dvali, R.~Kallosh and A.~Van Proeyen,
  ``Fayet-Iliopoulos terms in supergravity and cosmology,''
  Class.\ Quant.\ Grav.\  {\bf 21}, 3137 (2004)
  [arXiv:hep-th/0402046].
  %%CITATION = CQGRD,21,3137;%%
}

%\FerraraDH
\lref\FerraraDH{
  S.~Ferrara, L.~Girardello, T.~Kugo and A.~Van Proeyen,
  ``Relation Between Different Auxiliary Field Formulations Of N=1 Supergravity
  Coupled To Matter,''
  Nucl.\ Phys.\  B {\bf 223}, 191 (1983).
  %%CITATION = NUPHA,B223,191;%%
}

%\Wein\bergUV
\lref\WeinbergUV{
  S.~Weinberg,
  ``Non-renormalization theorems in non-renormalizable theories,''
  Phys.\ Rev.\ Lett.\  {\bf 80}, 3702 (1998)
  [arXiv:hep-th/9803099].
  %%CITATION = PRLTA,80,3702;%%
}

%\FerraraPZ
\lref\FerraraPZ{
  S.~Ferrara and B.~Zumino,
  ``Transformation Properties Of The Supercurrent,''
  Nucl.\ Phys.\  B {\bf 87}, 207 (1975).
  %%CITATION = NUPHA,B87,207;%%
}

%\DineTA
\lref\DineTA{
  M.~Dine,
  ``Fields, Strings and Duality: TASI 96,''
 eds. C.~Efthimiou and B. Greene (World Scientific, Singapore, 1997).
   }

%\WittenBZ
\lref\WittenBZ{
  E.~Witten,
  ``New Issues In Manifolds Of SU(3) Holonomy,''
  Nucl.\ Phys.\  B {\bf 268}, 79 (1986).
  %%CITATION = NUPHA,B268,79;%%
}

%\SeibergVC
\lref\SeibergVC{
  N.~Seiberg,
 ``Naturalness Versus Supersymmetric Non-renormalization Theorems,''
  Phys.\ Lett.\  B {\bf 318}, 469 (1993)
  [arXiv:hep-ph/9309335].
  %%CITATION = PHLTA,B318,469;%%
}

%\O'RaifeartaighPR
\lref\ORaifeartaighPR{
  L.~O'Raifeartaigh,
  ``Spontaneous Symmetry Breaking For Chiral Scalar Superfields,''
  Nucl.\ Phys.\  B {\bf 96}, 331 (1975).
  %%CITATION = NUPHA,B96,331;%%
}

%\FayetJB
\lref\FayetJB{
  P.~Fayet and J.~Iliopoulos,
  ``Spontaneously Broken Supergauge Symmetries and Goldstone Spinors,''
  Phys.\ Lett.\  B {\bf 51}, 461 (1974).
  %%CITATION = PHLTA,B51,461;%%
}

%\GreenSG
\lref\GreenSG{
  M.~B.~Green and J.~H.~Schwarz,
  ``Anomaly Cancellation In Supersymmetric D=10 Gauge Theory And Superstring
  Theory,''
  Phys.\ Lett.\  B {\bf 149}, 117 (1984).
  %%CITATION = PHLTA,B149,117;%%
}

%\ChamseddineGB
\lref\ChamseddineGB{
  A.~H.~Chamseddine and H.~K.~Dreiner,
  ``Anomaly Free Gauged R Symmetry In Local Supersymmetry,''
  Nucl.\ Phys.\  B {\bf 458}, 65 (1996)
  [arXiv:hep-ph/9504337].
  %%CITATION = NUPHA,B458,65;%%
}

%\FerraraMV
\lref\FerraraMV{
  S.~Ferrara and B.~Zumino,
  ``Structure Of Conformal Supergravity,''
  Nucl.\ Phys.\  B {\bf 134}, 301 (1978).
  %%CITATION = NUPHA,B134,301;%%
}

%\WittenYA
\lref\WittenYA{
  E.~Witten,
  ``SL(2,Z) action on three-dimensional conformal field theories with Abelian
  symmetry,''
  arXiv:hep-th/0307041.
  %%CITATION = HEP-TH/0307041;%%
}

%\CastanoCI
\lref\CastanoCI{
  D.~J.~Castano, D.~Z.~Freedman and C.~Manuel,
  ``Consequences of supergravity with gauged U(1)-R symmetry,''
  Nucl.\ Phys.\  B {\bf 461}, 50 (1996)
  [arXiv:hep-ph/9507397].
  %%CITATION = NUPHA,B461,50;%%
}

%\WittenHU
\lref\WittenHU{
  E.~Witten and J.~Bagger,
  ``Quantization Of Newton's Constant In Certain Supergravity Theories,''
  Phys.\ Lett.\  B {\bf 115}, 202 (1982).
  %%CITATION = PHLTA,B115,202;%%
}

%\BaggerFN
\lref\BaggerFN{
  J.~Bagger and E.~Witten,
  ``The Gauge Invariant Supersymmetric Nonlinear Sigma Model,''
  Phys.\ Lett.\  B {\bf 118}, 103 (1982).
  %%CITATION = PHLTA,B118,103;%%
}
%\FischlerAB
\lref\FischlerAB{ W.~Fischler and L.~Susskind,
  ``Holography and cosmology,''
  arXiv:hep-th/9806039.
  %%CITATION = HEP-TH/9806039;%%
}
%\BoussoXY
\lref\BoussoXY{
  R.~Bousso,
  ``A Covariant Entropy Conjecture,''
  JHEP {\bf 9907}, 004 (1999)
  [arXiv:hep-th/9905177].
  %%CITATION = JHEPA,9907,004;%%
}

%\GirardelloWZ
\lref\GirardelloWZ{
  L.~Girardello and M.~T.~Grisaru,
  ``Soft Breaking Of Supersymmetry,''
  Nucl.\ Phys.\  B {\bf 194}, 65 (1982).
  %%CITATION = NUPHA,B194,65;%%
}

%\AffleckVC
\lref\AffleckVC{
  I.~Affleck, M.~Dine and N.~Seiberg,
  ``Dynamical Supersymmetry Breaking In Chiral Theories,''
  Phys.\ Lett.\  B {\bf 137}, 187 (1984).
  %%CITATION = PHLTA,B137,187;%%
}

%\WessCP
\lref\WessCP{
  J.~Wess and J.~Bagger,
  ``Supersymmetry and supergravity,''
%\href{/spires/find/hep/www?irn=5426545}{SPIRES entry}
{\it  Princeton, USA: Univ. Pr. (1992) 259 p}
}

%\BrignoleSK
\lref\BrignoleSK{
  A.~Brignole, F.~Feruglio and F.~Zwirner,
  ``Signals of a superlight gravitino at e+ e- colliders when the other
  superparticles are heavy,''
  Nucl.\ Phys.\  B {\bf 516}, 13 (1998)
  [Erratum-ibid.\  B {\bf 555}, 653 (1999)]
  [arXiv:hep-ph/9711516].
  %%CITATION = NUPHA,B516,13;%%
}

%\DineXI
\lref\DineXI{
  M.~Dine, N.~Seiberg and S.~Thomas,
  ``Higgs Physics as a Window Beyond the MSSM (BMSSM),''
  Phys.\ Rev.\  D {\bf 76}, 095004 (2007)
  [arXiv:0707.0005 [hep-ph]].
  %%CITATION = PHRVA,D76,095004;%%
}

%\KomargodskiPC
\lref\KomargodskiPC{
  Z.~Komargodski and N.~Seiberg,
  ``Comments on the Fayet-Iliopoulos Term in Field Theory and Supergravity,''
  JHEP {\bf 0906}, 007 (2009)
  [arXiv:0904.1159 [hep-th]].
  %%CITATION = JHEPA,0906,007;%%
}

%\AffleckVC
\lref\AffleckVC{
  I.~Affleck, M.~Dine and N.~Seiberg,
  ``Dynamical Supersymmetry Breaking In Chiral Theories,''
  Phys.\ Lett.\  B {\bf 137}, 187 (1984).
  %%CITATION = PHLTA,B137,187;%%
}

%\ClarkBG
\lref\ClarkBG{
  T.~E.~Clark and S.~T.~Love,
  ``The Supercurrent in supersymmetric field theories,''
  Int.\ J.\ Mod.\ Phys.\  A {\bf 11}, 2807 (1996)
  [arXiv:hep-th/9506145].
  %%CITATION = IMPAE,A11,2807;%%
}

%\BrignoleFN
\lref\BrignoleFN{
  A.~Brignole, F.~Feruglio and F.~Zwirner,
  ``Aspects of spontaneously broken N = 1 global supersymmetry in the  presence
  of gauge interactions,''
  Nucl.\ Phys.\  B {\bf 501}, 332 (1997)
  [arXiv:hep-ph/9703286].
  %%CITATION = NUPHA,B501,332;%%
}

%\AffleckMF
\lref\AffleckMF{
  I.~Affleck, M.~Dine and N.~Seiberg,
  ``Exponential Hierarchy From Dynamical Supersymmetry Breaking,''
  Phys.\ Lett.\  B {\bf 140}, 59 (1984).
  %%CITATION = PHLTA,B140,59;%%
}

%\AffleckXZ
\lref\AffleckXZ{
  I.~Affleck, M.~Dine and N.~Seiberg,
  ``Dynamical Supersymmetry Breaking In Four-Dimensions And Its
  Phenomenological Implications,''
  Nucl.\ Phys.\  B {\bf 256}, 557 (1985).
  %%CITATION = NUPHA,B256,557;%%
}

%\WeinbergKR
\lref\WeinbergKR{
  S.~Weinberg,
  ``The quantum theory of fields. Vol. 2: Modern applications,''
%\href{/spires/find/hep/www?irn=3763846}{SPIRES entry}
{\it  Cambridge, UK: Univ. Pr. (1996) 489 p}
}

%\UematsuRJ
\lref\UematsuRJ{
  T.~Uematsu and C.~K.~Zachos,
  ``Structure Of Phenomenological Lagrangians For Broken Supersymmetry,''
  Nucl.\ Phys.\  B {\bf 201}, 250 (1982).
  %%CITATION = NUPHA,B201,250;%%
}

%\ElitzurUV
\lref\ElitzurUV{
  S.~Elitzur, R.~B.~Pearson and J.~Shigemitsu,
  ``The Phase Structure Of Discrete Abelian Spin And Gauge Systems,''
  Phys.\ Rev.\  D {\bf 19}, 3698 (1979).
  %%CITATION = PHRVA,D19,3698;%%
}
%\UkawaYV
\lref\UkawaYV{
  A.~Ukawa, P.~Windey and A.~H.~Guth,
  ``Dual Variables For Lattice Gauge Theories And The Phase Structure Of Z(N)
  Systems,''
  Phys.\ Rev.\  D {\bf 21}, 1013 (1980).
  %%CITATION = PHRVA,D21,1013;%%
}

%\SchwingerNM
\lref\SchwingerNM{
  J.~S.~Schwinger,
  ``On gauge invariance and vacuum polarization,''
  Phys.\ Rev.\  {\bf 82}, 664 (1951).
  %%CITATION = PHRVA,82,664;%%
}

%\DineII
\lref\DineII{
  M.~Dine, R.~Kitano, A.~Morisse and Y.~Shirman,
  ``Moduli decays and gravitinos,''
  Phys.\ Rev.\  D {\bf 73}, 123518 (2006)
  [arXiv:hep-ph/0604140].
  %%CITATION = PHRVA,D73,123518;%%
}

%\WittenYC
\lref\WittenYC{
  E.~Witten,
  ``Phases of N = 2 theories in two dimensions,''
  Nucl.\ Phys.\  B {\bf 403}, 159 (1993)
  [arXiv:hep-th/9301042].
  %%CITATION = NUPHA,B403,159;%%
}

%\KachruEM
\lref\KachruEM{
  S.~Kachru, J.~McGreevy and P.~Svrcek,
  ``Bounds on masses of bulk fields in string compactifications,''
  JHEP {\bf 0604}, 023 (2006)
  [arXiv:hep-th/0601111].
  %%CITATION = JHEPA,0604,023;%%
}

%\KachruXP
\lref\KachruXP{
  S.~Kachru, L.~McAllister and R.~Sundrum,
  ``Sequestering in string theory,''
  JHEP {\bf 0710}, 013 (2007)
  [arXiv:hep-th/0703105].
  %%CITATION = JHEPA,0710,013;%%
}

%\RandallUK
\lref\RandallUK{
  L.~Randall and R.~Sundrum,
  ``Out of this world supersymmetry breaking,''
  Nucl.\ Phys.\  B {\bf 557}, 79 (1999)
  [arXiv:hep-th/9810155].
  %%CITATION = NUPHA,B557,79;%%
}

%\DienesTD
\lref\DienesTD{
  K.~R.~Dienes and B.~Thomas,
  ``On the Inconsistency of Fayet-Iliopoulos Terms in Supergravity Theories,''
  arXiv:0911.0677 [hep-th].
  %%CITATION = ARXIV:0911.0677;%%
}

%\ArkaniHamedMJ
\lref\ArkaniHamedMJ{
  N.~Arkani-Hamed and H.~Murayama,
  ``Holomorphy, rescaling anomalies and exact beta functions in  supersymmetric
  gauge theories,''
  JHEP {\bf 0006}, 030 (2000)
  [arXiv:hep-th/9707133].
  %%CITATION = JHEPA,0006,030;%%
}

%\BanksVH
\lref\BanksVH{
  T.~Banks, W.~Fischler, S.~H.~Shenker and L.~Susskind,
  ``M theory as a matrix model: A conjecture,''
  Phys.\ Rev.\  D {\bf 55}, 5112 (1997)
  [arXiv:hep-th/9610043].
  %%CITATION = PHRVA,D55,5112;%%
}

%\AharonyTI
\lref\AharonyTI{
  O.~Aharony, S.~S.~Gubser, J.~M.~Maldacena, H.~Ooguri and Y.~Oz,
  ``Large N field theories, string theory and gravity,''
  Phys.\ Rept.\  {\bf 323}, 183 (2000)
  [arXiv:hep-th/9905111].
  %%CITATION = PRPLC,323,183;%%
}

%\GrisaruYK
\lref\GrisaruYK{
  M.~T.~Grisaru, B.~Milewski and D.~Zanon,
  ``Supercurrents, Anomalies And The Adler-Bardeen Theorem,''
  Phys.\ Lett.\  B {\bf 157}, 174 (1985).
  %%CITATION = PHLTA,B157,174;%%
}

%\GrisaruIK
\lref\GrisaruIK{
  M.~T.~Grisaru, B.~Milewski and D.~Zanon,
  ``The Supercurrent And The Adler-Bardeen Theorem,''
  Nucl.\ Phys.\  B {\bf 266}, 589 (1986).
  %%CITATION = NUPHA,B266,589;%%
}

%\ShifmanZI
\lref\ShifmanZI{
  M.~A.~Shifman and A.~I.~Vainshtein,
 ``Solution of the Anomaly Puzzle in SUSY Gauge Theories and the Wilson
 Operator Expansion,''
  Nucl.\ Phys.\  B {\bf 277}, 456 (1986)
  [Sov.\ Phys.\ JETP {\bf 64}, 428 (1986\ ZETFA,91,723-744.1986)].
  %%CITATION = ZETFA,91,723;%%
}

%\WeinbergCR
\lref\WeinbergCR{
  S.~Weinberg,
  ``The quantum theory of fields.  Vol. 3: Supersymmetry,''
%\href{http://www.slac.stanford.edu/spires/find/hep/www?irn=4384008}{SPIRES entry}
{\it  Cambridge, UK: Univ. Pr. (2000) 419 p} }

%\StelleYE
\lref\StelleYE{
  K.~S.~Stelle and P.~C.~West,
  ``Minimal Auxiliary Fields For Supergravity,''
  Phys.\ Lett.\  B {\bf 74}, 330 (1978).
  %%CITATION = PHLTA,B74,330;%%
}

%\FerraraEM
\lref\FerraraEM{
  S.~Ferrara and P.~van Nieuwenhuizen,
  ``The Auxiliary Fields Of Supergravity,''
  Phys.\ Lett.\  B {\bf 74}, 333 (1978).
  %%CITATION = PHLTA,B74,333;%%
}

%\SohniusTP
\lref\SohniusTP{
  M.~F.~Sohnius and P.~C.~West,
  ``An Alternative Minimal Off-Shell Version Of N=1 Supergravity,''
  Phys.\ Lett.\  B {\bf 105}, 353 (1981).
  %%CITATION = PHLTA,B105,353;%%
}

%\GirardiVQ
\lref\GirardiVQ{
  G.~Girardi, R.~Grimm, M.~Muller and J.~Wess,
  ``Antisymmetric Tensor Gauge Potential In Curved Superspace And A (16+16)
  Supergravity Multiplet,''
  Phys.\ Lett.\  B {\bf 147}, 81 (1984).
  %%CITATION = PHLTA,B147,81;%%
}

%\LangXK
\lref\LangXK{
  W.~Lang, J.~Louis and B.~A.~Ovrut,
  ``(16+16) Supergravity Coupled To Matter: The Low-Energy Limit Of The
  Superstring,''
  Phys.\ Lett.\  B {\bf 158}, 40 (1985).
  %%CITATION = PHLTA,B158,40;%%
}

%\SiegelSV
\lref\SiegelSV{
  W.~Siegel,
  ``16/16 Supergravity,''
  Class.\ Quant.\ Grav.\  {\bf 3}, L47 (1986).
  %%CITATION = CQGRD,3,L47;%%
}

%\HuangTN
\lref\HuangTN{
  X.~Huang and L.~Parker,
  ``Clarifying Some Remaining Questions in the Anomaly Puzzle in ${\cal N} = 1$
  Supersymmetric Yang-Mills Theory,''
  arXiv:1001.2364 [hep-th].
  %%CITATION = ARXIV:1001.2364;%%
}

%\DvaliZH
\lref\DvaliZH{
  G.~Dvali, R.~Kallosh and A.~Van Proeyen,
  ``D-term strings,''
  JHEP {\bf 0401}, 035 (2004)
  [arXiv:hep-th/0312005].
  %%CITATION = JHEPA,0401,035;%%
}

%\KachruAW
\lref\KachruAW{
  S.~Kachru, R.~Kallosh, A.~D.~Linde and S.~P.~Trivedi,
  ``De Sitter vacua in string theory,''
  Phys.\ Rev.\  D {\bf 68}, 046005 (2003)
  [arXiv:hep-th/0301240].
  %%CITATION = PHRVA,D68,046005;%%
}

%\DouglasES
\lref\DouglasES{
  M.~R.~Douglas and S.~Kachru,
  ``Flux compactification,''
  Rev.\ Mod.\ Phys.\  {\bf 79}, 733 (2007)
  [arXiv:hep-th/0610102].
  %%CITATION = RMPHA,79,733;%%
}

%\RandallUK
\lref\RandallUK{
  L.~Randall and R.~Sundrum,
  ``Out of this world supersymmetry breaking,''
  Nucl.\ Phys.\  B {\bf 557}, 79 (1999)
  [arXiv:hep-th/9810155].
  %%CITATION = NUPHA,B557,79;%%
}

%\NovikovUC
\lref\NovikovUC{
  V.~A.~Novikov, M.~A.~Shifman, A.~I.~Vainshtein and V.~I.~Zakharov,
  ``Exact Gell-Mann-Low Function Of Supersymmetric Yang-Mills Theories From
  %Instanton Calculus,''
  Nucl.\ Phys.\  B {\bf 229}, 381 (1983).
  %%CITATION = NUPHA,B229,381;%%
}

%\NovikovIC
\lref\NovikovIC{
  V.~A.~Novikov, M.~A.~Shifman, A.~I.~Vainshtein and V.~I.~Zakharov,
  ``Supersymmetric instanton calculus: Gauge theories with matter,''
  Nucl.\ Phys.\  B {\bf 260}, 157 (1985)
  [Yad.\ Fiz.\  {\bf 42}, 1499 (1985)].
  %%CITATION = YAFIA,42,1499;%%
}

%\NovikovRD
\lref\NovikovRD{
  V.~A.~Novikov, M.~A.~Shifman, A.~I.~Vainshtein and V.~I.~Zakharov,
  ``Beta Function In Supersymmetric Gauge Theories: Instantons Versus
  Traditional Approach,''
  Phys.\ Lett.\  B {\bf 166}, 329 (1986)
  [Sov.\ J.\ Nucl.\ Phys.\  {\bf 43}, 294.1986\ YAFIA,43,459 (1986\ YAFIA,43,459-464.1986)].
  %%CITATION = YAFIA,43,459;%%
}

%\JonesIP
\lref\JonesIP{
  D.~R.~T.~Jones,
  ``More On The Axial Anomaly In Supersymmetric Yang-Mills Theory,''
  Phys.\ Lett.\  B {\bf 123}, 45 (1983).
  %%CITATION = PHLTA,B123,45;%%
}

%\JonesMI
\lref\JonesMI{
  D.~R.~T.~Jones, L.~Mezincescu and P.~C.~West,
  ``Anomalous Dimensions, Supersymmetry And The Adler-Bardeen Theorem,''
  Phys.\ Lett.\  B {\bf 151}, 219 (1985).
  %%CITATION = PHLTA,B151,219;%%
}

%\NovikovMF
\lref\NovikovMF{
  V.~A.~Novikov, M.~A.~Shifman, A.~I.~Vainshtein and V.~I.~Zakharov,
  ``Supersymmetric Extension Of The Adler-Bardeen Theorem,''
  Phys.\ Lett.\  B {\bf 157}, 169 (1985).
  %%CITATION = PHLTA,B157,169;%%
}

%\ArkaniHamedUT
\lref\ArkaniHamedUT{
  N.~Arkani-Hamed and H.~Murayama,
  ``Renormalization group invariance of exact results in supersymmetric  gauge
  theories,''
  Phys.\ Rev.\  D {\bf 57}, 6638 (1998)
  [arXiv:hep-th/9705189].
  %%CITATION = PHRVA,D57,6638;%%
}

%\ShifmanZI
\lref\ShifmanZI{
  M.~A.~Shifman and A.~I.~Vainshtein,
  ``Solution of the Anomaly Puzzle in SUSY Gauge Theories and the Wilson
  Operator Expansion,''
  Nucl.\ Phys.\  B {\bf 277}, 456 (1986)
  [Sov.\ Phys.\ JETP {\bf 64}, 428 (1986\ ZETFA,91,723-744.1986)].
  %%CITATION = ZETFA,91,723;%%
}

%\GatesAZ
\lref\GatesAZ{
  S.~J.~J.~Gates and W.~Siegel,
  ``Variant Superfield Representations,''
  Nucl.\ Phys.\  B {\bf 187}, 389 (1981).
  %%CITATION = NUPHA,B187,389;%%
}

%\FerraraDH
\lref\FerraraDH{
  S.~Ferrara, L.~Girardello, T.~Kugo and A.~Van Proeyen,
  ``Relation Between Different Auxiliary Field Formulations Of N=1 Supergravity
  Coupled To Matter,''
  Nucl.\ Phys.\  B {\bf 223}, 191 (1983).
  %%CITATION = NUPHA,B223,191;%%
}

%\AkulovCK
\lref\AkulovCK{
  V.~P.~Akulov, D.~V.~Volkov and V.~A.~Soroka,
  ``On The General Covariant Theory Of Calibrating Poles In Superspace,''
  Theor.\ Math.\ Phys.\  {\bf 31}, 285 (1977)
  [Teor.\ Mat.\ Fiz.\  {\bf 31}, 12 (1977)].
  %%CITATION = TMFZA,31,12;%%
}

%\GatesCZ
\lref\GatesCZ{
  S.~J.~J.~Gates, S.~M.~Kuzenko and J.~Phillips,
  ``The off-shell (3/2,2) supermultiplets revisited,''
  Phys.\ Lett.\  B {\bf 576}, 97 (2003)
  [arXiv:hep-th/0306288].
  %%CITATION = PHLTA,B576,97;%%
}

%\HartmanPB
\lref\HartmanPB{
  T.~Hartman, K.~Murata, T.~Nishioka and A.~Strominger,
  ``CFT Duals for Extreme Black Holes,''
  JHEP {\bf 0904}, 019 (2009)
  [arXiv:0811.4393 [hep-th]].
  %%CITATION = JHEPA,0904,019;%%
}

%\WittenTW
\lref\WittenTW{
  E.~Witten,
  ``Global Aspects Of Current Algebra,''
  Nucl.\ Phys.\  B {\bf 223}, 422 (1983).
  %%CITATION = NUPHA,B223,422;%%
}

%\SkyrmeVQ
\lref\SkyrmeVQ{
  T.~H.~R.~Skyrme,
  ``A Nonlinear field theory,''
  Proc.\ Roy.\ Soc.\ Lond.\  A {\bf 260}, 127 (1961).
  %%CITATION = PRSLA,A260,127;%%
}

%\FinkelsteinHY
\lref\FinkelsteinHY{
  D.~Finkelstein and J.~Rubinstein,
  ``Connection between spin, statistics, and kinks,''
  J.\ Math.\ Phys.\  {\bf 9}, 1762 (1968).
  %%CITATION = JMAPA,9,1762;%%
}

%\WittenTX
\lref\WittenTX{
  E.~Witten,
  ``Current Algebra, Baryons, And Quark Confinement,''
  Nucl.\ Phys.\  B {\bf 223}, 433 (1983).
  %%CITATION = NUPHA,B223,433;%%
}

%\SeibergQD
\lref\SeibergQD{
  N.~Seiberg,
  ``Modifying the Sum Over Topological Sectors and Constraints on
  %Supergravity,''
  arXiv:1005.0002 [hep-th].
  %%CITATION = ARXIV:1005.0002;%%
}

%\FradkinJQ
\lref\FradkinJQ{
  E.~S.~Fradkin and M.~A.~Vasiliev,
  ``S Matrix For Theories That Admit Closure Of The Algebra With The Aid Of
  Auxiliary Fields: The Auxiliary Fields In Supergravity,''
  Lett.\ Nuovo Cim.\  {\bf 22}, 651 (1978).
  %%CITATION = NCLTA,22,651;%%
}

%\GatesNR
\lref\GatesNR{
  S.~J.~Gates, M.~T.~Grisaru, M.~Rocek and W.~Siegel,
  ``Superspace, or one thousand and one lessons in supersymmetry,''
  Front.\ Phys.\  {\bf 58}, 1 (1983)
  [arXiv:hep-th/0108200].
  %%CITATION = FRPHA,58,1;%%
}

%\WeinbergCR
\lref\WeinbergCR{
  S.~Weinberg,
  ``The quantum theory of fields.  Vol. 3: Supersymmetry,''
%\href{http://www.slac.stanford.edu/spires/find/hep/www?irn=4384008}{SPIRES entry}
{\it  Cambridge, UK: Univ. Pr. (2000) 419 p}
}

%\DineVU
\lref\DineVU{
  M.~Dine, P.~Y.~Huet and N.~Seiberg,
  ``Large and Small Radius in String Theory,''
  Nucl.\ Phys.\  B {\bf 322}, 301 (1989).
  %%CITATION = NUPHA,B322,301;%%
}

%\WeinbergUV
\lref\WeinbergUV{
  S.~Weinberg,
  ``Non-renormalization theorems in non-renormalizable theories,''
  Phys.\ Rev.\ Lett.\  {\bf 80}, 3702 (1998)
  [arXiv:hep-th/9803099].
  %%CITATION = PRLTA,80,3702;%%
}

%\FerraraPZ
\lref\FerraraPZ{
  S.~Ferrara and B.~Zumino,
  ``Transformation Properties Of The Supercurrent,''
  Nucl.\ Phys.\  B {\bf 87}, 207 (1975).
  %%CITATION = NUPHA,B87,207;%%
}

%\IntriligatorCP
\lref\IntriligatorCP{
  K.~A.~Intriligator and N.~Seiberg,
  ``Lectures on Supersymmetry Breaking,''
  Class.\ Quant.\ Grav.\  {\bf 24}, S741 (2007)
  [arXiv:hep-ph/0702069].
  %%CITATION = CQGRD,24,S741;%%
}

%\DineTA
\lref\DineTA{
  M.~Dine,
  ``Fields, Strings and Duality: TASI 96,''
 eds. C.~Efthimiou and B. Greene (World Scientific, Singapore, 1997).
   }

%\FerraraPZ
\lref\FerraraPZ{
  S.~Ferrara and B.~Zumino,
  ``Transformation Properties Of The Supercurrent,''
  Nucl.\ Phys.\  B {\bf 87}, 207 (1975).
  %%CITATION = NUPHA,B87,207;%%
}

%\WittenBZ
\lref\WittenBZ{
  E.~Witten,
  ``New Issues In Manifolds Of SU(3) Holonomy,''
  Nucl.\ Phys.\  B {\bf 268}, 79 (1986).
  %%CITATION = NUPHA,B268,79;%%
}

%\HorowitzNG
\lref\HorowitzNG{
  G.~T.~Horowitz,
  ``Exactly Soluble Diffeomorphism Invariant Theories,''
  Commun.\ Math.\ Phys.\  {\bf 125}, 417 (1989).
  %%CITATION = CMPHA,125,417;%%
}

%\SeibergVC
\lref\SeibergVC{
  N.~Seiberg,
 ``Naturalness Versus Supersymmetric Non-renormalization Theorems,''
  Phys.\ Lett.\  B {\bf 318}, 469 (1993)
  [arXiv:hep-ph/9309335].
  %%CITATION = PHLTA,B318,469;%%
}

%\O'RaifeartaighPR
\lref\ORaifeartaighPR{
  L.~O'Raifeartaigh,
  ``Spontaneous Symmetry Breaking For Chiral Scalar Superfields,''
  Nucl.\ Phys.\  B {\bf 96}, 331 (1975).
  %%CITATION = NUPHA,B96,331;%%
}

%\FayetJB
\lref\FayetJB{
  P.~Fayet and J.~Iliopoulos,
  ``Spontaneously Broken Supergauge Symmetries and Goldstone Spinors,''
  Phys.\ Lett.\  B {\bf 51}, 461 (1974).
  %%CITATION = PHLTA,B51,461;%%
}

%\GreenSG
\lref\GreenSG{
  M.~B.~Green and J.~H.~Schwarz,
  ``Anomaly Cancellation In Supersymmetric D=10 Gauge Theory And Superstring
  Theory,''
  Phys.\ Lett.\  B {\bf 149}, 117 (1984).
  %%CITATION = PHLTA,B149,117;%%
}

%\PantevRH
\lref\PantevRH{
  T.~Pantev and E.~Sharpe,
  ``Notes on gauging noneffective group actions,''
  arXiv:hep-th/0502027.
  %%CITATION = HEP-TH/0502027;%%
}

%\PantevZS
\lref\PantevZS{
  T.~Pantev and E.~Sharpe,
  ``GLSM's for gerbes (and other toric stacks),''
  Adv.\ Theor.\ Math.\ Phys.\  {\bf 10}, 77 (2006)
  [arXiv:hep-th/0502053].
  %%CITATION = 00203,10,77;%%
}

%\ChamseddineGB
\lref\ChamseddineGB{
  A.~H.~Chamseddine and H.~K.~Dreiner,
  ``Anomaly Free Gauged R Symmetry In Local Supersymmetry,''
  Nucl.\ Phys.\  B {\bf 458}, 65 (1996)
  [arXiv:hep-ph/9504337].
  %%CITATION = NUPHA,B458,65;%%
}

%\CastanoCI
\lref\CastanoCI{
  D.~J.~Castano, D.~Z.~Freedman and C.~Manuel,
  ``Consequences of supergravity with gauged U(1)-R symmetry,''
  Nucl.\ Phys.\  B {\bf 461}, 50 (1996)
  [arXiv:hep-ph/9507397].
  %%CITATION = NUPHA,B461,50;%%
}

%\WittenHU
\lref\WittenHU{
  E.~Witten and J.~Bagger,
  ``Quantization Of Newton's Constant In Certain Supergravity Theories,''
  Phys.\ Lett.\  B {\bf 115}, 202 (1982).
  %%CITATION = PHLTA,B115,202;%%
}

%\BaggerFN
\lref\BaggerFN{
  J.~Bagger and E.~Witten,
  ``The Gauge Invariant Supersymmetric Nonlinear Sigma Model,''
  Phys.\ Lett.\  B {\bf 118}, 103 (1982).
  %%CITATION = PHLTA,B118,103;%%
}

%\DvaliXE
\lref\DvaliXE{
  G.~R.~Dvali and M.~A.~Shifman,
  ``Domain walls in strongly coupled theories,''
  Phys.\ Lett.\  B {\bf 396}, 64 (1997)
  [Erratum-ibid.\  B {\bf 407}, 452 (1997)]
  [arXiv:hep-th/9612128].
  %%CITATION = PHLTA,B396,64;%%

}

%\BanksMB
\lref\BanksMB{
  T.~Banks, M.~Dine and N.~Seiberg,
  ``Irrational axions as a solution of the strong CP problem in an eternal
  universe,''
  Phys.\ Lett.\  B {\bf 273}, 105 (1991)
  [arXiv:hep-th/9109040].
  %%CITATION = PHLTA,B273,105;%%
}

%\KomargodskiRB
\lref\KomargodskiRB{
  Z.~Komargodski and N.~Seiberg,
  ``Comments on Supercurrent Multiplets, Supersymmetric Field Theories and
  Supergravity,''
  arXiv:1002.2228 [hep-th].
  %%CITATION = ARXIV:1002.2228;%%
}

%\KuzenkoAM
\lref\KuzenkoAM{
  S.~M.~Kuzenko,
  ``Variant supercurrent multiplets,''
  arXiv:1002.4932 [hep-th].
  %%CITATION = ARXIV:1002.4932;%%
}

%\DienesTD
\lref\DienesTD{
  K.~R.~Dienes and B.~Thomas,
  ``On the Inconsistency of Fayet-Iliopoulos Terms in Supergravity Theories,''
  arXiv:0911.0677 [hep-th].
  %%CITATION = ARXIV:0911.0677;%%
}

%\KuzenkoYM
\lref\KuzenkoYM{
  S.~M.~Kuzenko,
  ``The Fayet-Iliopoulos term and nonlinear self-duality,''
  arXiv:0911.5190 [hep-th].
  %%CITATION = ARXIV:0911.5190;%%
}

%\WittenBZ
\lref\WittenBZ{
  E.~Witten,
  ``New Issues In Manifolds Of SU(3) Holonomy,''
  Nucl.\ Phys.\  B {\bf 268}, 79 (1986).
  %%CITATION = NUPHA,B268,79;%%
}

%\KugoFS
\lref\KugoFS{
  T.~Kugo and T.~T.~Yanagida,
  ``Coupling Supersymmetric Nonlinear Sigma Models to Supergravity,''
  arXiv:1003.5985 [hep-th].
  %%CITATION = ARXIV:1003.5985;%%
}

\lref\arkani{N.~Arkani-Hamed,  }

 \lref\emparan{R.~Emparan, G.~Horowitz, R.~Myers, }

\def\S{{\bf S}}

\def\Z{{\bf Z}}
\def\R{{\bf R}}

\def\K3{{\bf K3}}
\def\journal#1&#2(#3){\unskip, \sl #1\ \bf #2 \rm(19#3) }
\def\andjournal#1&#2(#3){\sl #1~\bf #2 \rm (19#3) }

\def\bar{\overline}

\def\tilde{\widetilde}

\def\frac#1#2{{#1\over#2}}

\def\inbar{\,\vrule height1.5ex width.4pt depth0pt}
\def\IC{\relax\hbox{$\inbar\kern-.3em{\rm C}$}}
\def\IR{\relax{\rm I\kern-.18em R}}
\def\IP{\relax{\rm I\kern-.18em P}}
\def\Z{{\bf Z}}

%
%%%%%%%%%%%%%%%%%%%%%%%%%%%%%%%%%%%%
%

%
\catcode`\@=11
\def\slash#1{\mathord{\mathpalette\c@ncel{#1}}}
\overfullrule=0pt

\def\underrel#1\over#2{\mathrel{\mathop{\kern\z@#1}\limits_{#2}}}

\catcode`\@=12

%%%%%%%%%%%%%%%%%%%%%%%%%%%%%%%%%%%%%%%%%%%%%%%%%%%%%%%%%%%%%%

%

\def\mod{{\rm mod}}

\def\exp{{\rm exp}}

\def\unit{\relax{\rm 1\kern-.26em I}}
\def\nada{\relax{\rm 0\kern-.30em l}}
\def\tilde{\widetilde}

\def\alphadot{{\dot \alpha}}

% \draftmode

%\def\Omega{\rho,\sigma,\nu  }

\def\mod{{\rm mod}}

\def\CP{{\cal P}}
%% MACROS
\noblackbox
\def\IL{\relax{\rm I\kern-.18em L}}
\def\IH{\relax{\rm I\kern-.18em H}}
\def\IR{\relax{\rm I\kern-.18em R}}
\def\IC{\relax\hbox{$\inbar\kern-.3em{\rm C}$}}
\def\IZ{\relax\ifmmode\mathchoice
{\hbox{\cmss Z\kern-.4em Z}}{\hbox{\cmss Z\kern-.4em Z}} {\lower.9pt\hbox{\cmsss Z\kern-.4em Z}}
{\lower1.2pt\hbox{\cmsss Z\kern-.4em Z}}\else{\cmss Z\kern-.4em Z}\fi}

\def\CN {{\cal N}}
\def\CR {{\cal R}}

\def\CJ {{\cal J}}
\def\CP {{\cal P }}
\def\CL {{\cal L}}
\def\CV {{\cal V}}
\def\CO {{\cal O}}

\def\CH {{\cal H}}

\def\CB {{\cal B}}

\def\CS {{\cal S}}

%% MORE MACROS

\def\CN {{\cal N}}

\def\CO {{\cal O}}

\def\CP {{\cal P }}

\def\CV{{\cal V }}

\def\CS {{\cal S }}

\font\manual=manfnt \def\dbend{\lower3.5pt\hbox{\manual\char127}}

\def\IZ{\relax\ifmmode\mathchoice
{\hbox{\cmss Z\kern-.4em Z}}{\hbox{\cmss Z\kern-.4em Z}} {\lower.9pt\hbox{\cmsss Z\kern-.4em Z}}
{\lower1.2pt\hbox{\cmsss Z\kern-.4em Z}}\else{\cmss Z\kern-.4em Z}\fi}

\def\alphadot{{\dot \alpha}}

\def\bar{\overline}
\def\CS{{\cal S}}
\def\CH{{\cal H}}

\def\rt2{\sqrt{2}}
\def\irt2{{1\over\sqrt{2}}}

%  \slashchar puts a slash through a character to represent contraction
%  with Dirac matrices. Use \not instead for negation of relations, and use
%  \hbar for hbar.
\def\slashchar#1{\setbox0=\hbox{$#1$}           % set a box for #1
   \dimen0=\wd0                                 % and get its size
   \setbox1=\hbox{/} \dimen1=\wd1               % get size of /
   \ifdim\dimen0>\dimen1                        % #1 is bigger
      \rlap{\hbox to \dimen0{\hfil/\hfil}}      % so center / in box
      #1                                        % and print #1
   \else                                        % / is bigger
      \rlap{\hbox to \dimen1{\hfil$#1$\hfil}}   % so center #1
      /                                         % and print /
   \fi}

\def\foursqr#1#2{{\vcenter{\vbox{
    \hrule height.#2pt
    \hbox{\vrule width.#2pt height#1pt \kern#1pt
    \vrule width.#2pt}
    \hrule height.#2pt
    \hrule height.#2pt
    \hbox{\vrule width.#2pt height#1pt \kern#1pt
    \vrule width.#2pt}
    \hrule height.#2pt
        \hrule height.#2pt
    \hbox{\vrule width.#2pt height#1pt \kern#1pt
    \vrule width.#2pt}
    \hrule height.#2pt
        \hrule height.#2pt
    \hbox{\vrule width.#2pt height#1pt \kern#1pt
    \vrule width.#2pt}
    \hrule height.#2pt}}}}
\def\psqr#1#2{{\vcenter{\vbox{\hrule height.#2pt
    \hbox{\vrule width.#2pt height#1pt \kern#1pt
    \vrule width.#2pt}
    \hrule height.#2pt \hrule height.#2pt
    \hbox{\vrule width.#2pt height#1pt \kern#1pt
    \vrule width.#2pt}
    \hrule height.#2pt}}}}
\def\sqr#1#2{{\vcenter{\vbox{\hrule height.#2pt
    \hbox{\vrule width.#2pt height#1pt \kern#1pt
    \vrule width.#2pt}
    \hrule height.#2pt}}}}

\def\figin{\epsfcheck\figin}\def\figins{\epsfcheck\figins}
\def\epsfcheck{\ifx\epsfbox\UnDeFiNeD
\message{(NO epsf.tex, FIGURES WILL BE IGNORED)}
\gdef\figin##1{\vskip2in}\gdef\figins##1{\hskip.5in}% blank space instead
\else\message{(FIGURES WILL BE INCLUDED)}%
\gdef\figin##1{##1}\gdef\figins##1{##1}\fi}
\def\DefWarn#1{}
\def\figinsert{\goodbreak\midinsert}
\def\ifig#1#2#3{\DefWarn#1\xdef#1{fig.~\the\figno}
\writedef{#1\leftbracket fig.\noexpand~\the\figno}%
\figinsert\figin{\centerline{#3}}\medskip\centerline{\vbox{\baselineskip12pt \advance\hsize by
-1truein\noindent\footnotefont{\bf Fig.~\the\figno:\ } \it#2}}
\bigskip\endinsert\global\advance\figno by1}

%%%%%%%%%%%%%%%%%%%%%%%%%%%%%%%%%%%%%%%%%%%%%%%%%%%%%%%%%%%%%%
% new defs:

% FONTS

% fraktur

\newfam\frakfam
\font\teneufm=eufm10
\font\seveneufm=eufm7
\font\fiveeufm=eufm5
\textfont\frakfam=\teneufm
\scriptfont\frakfam=\seveneufm
\scriptscriptfont\frakfam=\fiveeufm

% black board bold

\def\bb{
\font\tenmsb=msbm10
\font\sevenmsb=msbm7
\font\fivemsb=msbm5
\textfont1=\tenmsb
\scriptfont1=\sevenmsb
\scriptscriptfont1=\fivemsb
}

%\newfam\msbfam
%\font\tenmsb=msbm10
%\font\sevenmsb=msbm7
%\font\fivemsb=msbm5
%\textfont\msbfam=\tenmsb
%\scriptfont\msbfam=\sevenmsb
%\scriptscriptfont\msbfam=\fivemsb
%\def\bb{\fam\msbfam \tenmsb}

% double stroke math

\newfam\dsromfam
\font\tendsrom=dsrom10
\textfont\dsromfam=\tendsrom
\def\ds{\fam\dsromfam \tendsrom}

% bold math italics

\newfam\mbffam
\font\tenmbf=cmmib10
\font\sevenmbf=cmmib7
\font\fivembf=cmmib5
\textfont\mbffam=\tenmbf
\scriptfont\mbffam=\sevenmbf
\scriptscriptfont\mbffam=\fivembf

% bold math cal

\newfam\mbfcalfam
\font\tenmbfcal=cmbsy10
\font\sevenmbfcal=cmbsy7
\font\fivembfcal=cmbsy5
\textfont\mbfcalfam=\tenmbfcal
\scriptfont\mbfcalfam=\sevenmbfcal
\scriptscriptfont\mbfcalfam=\fivembfcal

% math script

\newfam\mscrfam
\font\tenmscr=rsfs10
\font\sevenmscr=rsfs7
\font\fivemscr=rsfs5
\textfont\mscrfam=\tenmscr
\scriptfont\mscrfam=\sevenmscr
\scriptscriptfont\mscrfam=\fivemscr

% MACROS

% bras, kets, ...

% tilde, hat, bar, ...

\def\tilde{\widetilde}

\def\bar{\overline}
\def\b{\bar}
\def\bsq#1{{{\b{#1}}^{\lower 2.5pt\hbox{$\scriptstyle 2$}}}}
\def\bexp#1#2{{{\b{#1}}^{\lower 2.5pt\hbox{$\scriptstyle #2$}}}}
\def\dotexp#1#2{{{#1}^{\lower 2.5pt\hbox{$\scriptstyle #2$}}}}

% basic math

\def\rt2{\sqrt{2}}

\def\mod{{\rm mod}}

% bold greek characters

\font\tenbifull=cmmib10
\font\tenbimed=cmmib7
\font\tenbismall=cmmib5
\textfont9=\tenbifull \scriptfont9=\tenbimed
\scriptscriptfont9=\tenbismall

\mathchardef\bbGamma="7000
\mathchardef\bbDelta="7001
\mathchardef\bbPhi="7002
\mathchardef\bbAlpha="7003
\mathchardef\bbXi="7004
\mathchardef\bbPi="7005
\mathchardef\bbSigma="7006
\mathchardef\bbUpsilon="7007
\mathchardef\bbTheta="7008
\mathchardef\bbPsi="7009
\mathchardef\bbOmega="700A
\mathchardef\bbalpha="710B
\mathchardef\bbbeta="710C
\mathchardef\bbgamma="710D
\mathchardef\bbdelta="710E
\mathchardef\bbepsilon="710F
\mathchardef\bbzeta="7110
\mathchardef\bbeta="7111
\mathchardef\bbtheta="7112
\mathchardef\bbiota="7113
\mathchardef\bbkappa="7114
\mathchardef\bblambda="7115
\mathchardef\bbmu="7116
\mathchardef\bbnu="7117
\mathchardef\bbxi="7118
\mathchardef\bbpi="7119
\mathchardef\bbrho="711A
\mathchardef\bbsigma="711B
\mathchardef\bbtau="711C
\mathchardef\bbupsilon="711D
\mathchardef\bbphi="711E
\mathchardef\bbchi="711F
\mathchardef\bbpsi="7120
\mathchardef\bbomega="7121
\mathchardef\bbvarepsilon="7122
\mathchardef\bbvartheta="7123
\mathchardef\bbvarpi="7124
\mathchardef\bbvarrho="7125
\mathchardef\bbvarsigma="7126
\mathchardef\bbvarphi="7127

% dotted spinor indices

\def\alphadot{{\dot\alpha}}

% bared indices

% bared spinors

% capital cal letters

\def\CB{{\cal B}}

\def\CH{{\cal H}}

\def\CJ{{\cal J}}

\def\CL{{\cal L}}

\def\CN{{\cal N}}
\def\CO{{\cal O}}
\def\CP{{\cal P}}

\def\CR{{\cal R}}
\def\CS{{\cal S}}

\def\CV{{\cal V}}

% double stroke symbols: unit matrix, reals, complex, quaternions, integers, natural numbers

\def\1{{\ds 1}}
\def\R{\hbox{$\bb R$}}

\def\Z{\hbox{$\bb Z$}}
\def\S{\hbox{$\bb S$}}

\def\CP{\hbox{$\bb CP$}}

% miscellaneous objects

%%%%%%%%%%%%%%%%%%%%%%%%%%%%%%%%%%%%%%%%%%%%%%%%%%%
\Title{ } {\vbox{\centerline{Symmetries and Strings in}
 \centerline{}
 \centerline{Field Theory and Gravity}
}}
%\medskip

\centerline{\it Tom Banks${}^{1,2,3}$ and Nathan Seiberg${}^1$ }
\medskip
\centerline{${}^1$School of Natural Sciences}
\centerline{Institute for Advanced Study}
\centerline{Einstein Drive, Princeton, NJ 08540}
\medskip
\centerline{${}^2$New High Energy Theory Center}
\centerline{Department of Physics}
\centerline{Rutgers University}
\centerline{Piscataway, NJ 08854}
\medskip
\centerline{${}^3$Santa Cruz Institute for Particle Physics}
\centerline{University of California}
\centerline{
Santa Cruz, CA 95064}

%\smallskip

%\vglue .3cm

\bigskip
\noindent We discuss aspects of global and gauged symmetries in
quantum field theory and quantum gravity, focusing on discrete gauge
symmetries. An effective Lagrangian description of $\Z_p$ gauge
theories shows that they are associated with an emergent $\Z_p$
one-form (Kalb-Ramond) gauge symmetry.  This understanding leads us
to uncover new observables and new phenomena in nonlinear
$\sigma$-models.  It also allows us to expand on Polchinski's
classification of cosmic strings. We argue that in models of quantum
gravity, there are no global symmetries, all continuous gauge
symmetries are compact, and all charges allowed by Dirac
quantization are present in the spectrum. These conjectures are not
new, but we present them from a streamlined and unified perspective.
Finally, our discussion about string charges and symmetries leads to
a more physical and more complete understanding of recently found
consistency conditions of supergravity.

\Date{November 2010}

\newsec{Introduction and conclusions}

This work addresses several seemingly unrelated topics. We will
comment on some subtleties in gauge theories and in particular in
discrete (e.g.\ $\Z_p$) gauge theories.  We will explore the
classifications of strings in four dimensions. We will discuss some
conjectures about symmetries in quantum gravity, and we will present
some constraints on consistent supergravity theories.  Even though
these topics seem to be distinct we will see that they are related. Our improved understanding of the field
theory of discrete gauge symmetries, is crucial to a proper analysis
of gravitational theories.

A key focus of our work is a discussion of three conjectures about
quantum gravity\foot{We will concentrate on models with exactly four
non-compact dimensions, which are asymptotically flat, but we expect
that some version of our conjectures is true more generally.}. Some
of them are widely accepted as ``folk theorems.'' Others are known
to some experts (see e.g.\ \refs{\GarfinkleQJ\HorowitzCD-\HartmanPB}
and in particular \PolchinskiBQ\ and the earlier reference
\MisnerMT). We present new, streamlined arguments for them, and
explore their inter-relation and their implications for other
issues. The conjectures are:
\item{1.} There are no global symmetries.  Here we slightly extend the
known ``no global symmetries'' dictum to include also discrete
symmetry groups and higher brane charges.  In particular, we argue
that all stable branes are associated with gauge symmetries. For
example, stable strings in four dimensions are associated with a
continuous or discrete one-form gauge symmetry (a.k.a.\ Kalb-Ramond (KR)
gauge symmetry)\foot{Our conventions for higher form gauge symmetries and gauge fields follows that of branes.  A $q$-brane has a $q+1$-dimensional world-volume.  It couples to a $q+1$-form gauge field, which has a $q$-form gauge symmetry.  In this terminology ordinary gauge symmetry is a $0$-form gauge symmetry and a global symmetry is a ``$-1$-form gauge symmetry."  We will use this terminology both for continuous and for discrete gauge symmetries.}.
\item{2.} All continuous gauge groups, including Kalb-Ramond gauge symmetries, are compact.
This excludes for example, the possibility of a gauge group being $\R$ rather
than $U(1)$.
\item{3.} The Completeness Hypothesis: The spectrum of electric and magnetic
charges forms a complete set consistent with the Dirac quantization condition. By ``complete set'' we mean that every allowed charge is present in the spectrum.

Clearly, these conjectures are satisfied in all known examples of
string theory, i.e.\ in perturbative string theory, Matrix Theory,
and for strings in an asymptotically AdS background.

Our arguments in favor of the three conjectures show that the
distinction between observable operators and dynamical particles,
which pervades quantum field theory, disappears in models of quantum
gravity. In particular, Wilson and 't Hooft loops {\it must}
correspond to propagating particles, which might be stable charged
black holes.

Since our arguments thread together a number of themes and results,
which are at first sight quite unrelated, we end this introduction
with a reader's guide to the sections and their relationships.

Section 2: Here we review properties of gauge theories, many of
which are well known. We emphasize the analysis of these theories in
terms of their deep infrared (IR) behavior. We give a universal IR
Lagrangian for $\Z_p$ gauge theories in four dimensions.  This
involves a two-form and a one-form gauge potential, whose holonomies are valued in $U(1)$. The integer $p$ enters in the Lagrangian.  In sections 2 and 3 we will strive to find universal
properties of the long distance quantum field theory, as well as a
description of operators in the IR theory, which parameterize
possible objects in {\it any} UV completion of the theory. In some
more complete effective field theory at a higher scale, we may find
some of these objects as dynamical (perhaps solitonic) particles or
strings. The others remain as observables (Wilson lines {\it etc.}).
In section 4, we will argue that in models including gravity, all
the observables represent dynamical objects.

We also study a $\Z_p$ gauge theory coupled to $\sigma$-models, like
the $SO(3)$ model, with non-trivial two-cycles in their target
space. Here we use our improved understanding of the $\Z_p$ gauge
theories to shed new light on the constructions of \SeibergQD.  The
strings of these models might be a macroscopic manifestation of the
strings of a $\Z_p$ gauge theory. We point out the existence of
point, line and surface operators, which describe the violation of
topological symmetries of $\sigma$-models in a universal manner
independent of the UV completion of the model. We describe
applications of these operators  to violation of the string
conservation law of the $\sigma$-models referred to above. The
latter operators are analogs of magnetic monopoles for
Nielsen-Olesen strings. In Appendix B we note the application of
these operators to baryon number violation and the existence of
confining strings, in chiral Lagrangians derived from QCD like gauge
theories. Appendix C outlines how all of these questions arise in
lattice models.

Section 3: Here we review and refine Polchinski's classification of
cosmic strings \PolchinskiBG. Our general conclusion is that stable
strings are always coupled to a KR gauge field, which might be a
$\Z_p$ gauge field of the type described in section 2. In the latter
case $p$ strings can end at a point.  We also mention the
possibility of discrete non-Abelian strings. If the KR gauge group
is continuous, the corresponding $2$-form gauge field is massless and is dual to a scalar.  Here we argue that the theory should be supersymmetric in order to avoid a potential for that scalar.  Furthermore, in order to render the string tension finite, that scalar should be part of a moduli space of vacua and the latter should have a singularity around which the scalar winds. In addition, the boundary conditions should set the moduli to that singular value.  Otherwise, in the presence of gravity the strings cannot be viewed as excitations of the model. If one tries to create a large loop of string in
such a model, there will always be a radius above which one forms a
black hole instead\foot{One should be careful about applying many of
our conclusions, including this one, to phenomenology.  In many
cases, the instabilities we describe are exponential in the ratio of
the cosmic string tension scale and the Planck scale.}.

Section 4: Here we formulate the three interlocking conjectures about
symmetries in models of quantum gravity we mentioned above.  The arguments for these conjectures are
all based on black hole physics.  They are also valid in all {\it
known} consistent models of quantum gravity.

We argue that the covariant entropy bound (CEB) leads to an elegant
and simple way to formulate the principle of quantum gravity that
forbids global continuous symmetries. The same argument shows that
all continuous gauge groups are compact. We use the CEB and other
arguments from black hole physics to argue for the completeness
hypotheses and to show that all global cosmic string charges are
violated as well. Stable cosmic strings must be coupled to a gauge
field, which might be discrete. In section 3 we use these
observations to refine Polchinski's classification of cosmic
strings.

Section 5: We return to the origins of this paper
\refs{\KomargodskiPC,\KomargodskiRB,\SeibergQD}\ and explore the
consequences of all of these results for $\CN=1$ supergravity in
four dimensions.  These papers argued that certain rigid
supersymmetric theories cannot be coupled to supergravity in a
straightforward way. In particular, problems arise either when the
theory includes Fayet-Iliopoulos (FI) terms or when the target space
topology has a non-exact K\"ahler form.  We will give a more
physical interpretation of these results by emphasizing the role of
a certain $2$-form string current\foot{More generally, $q$-branes
can carry a conserved $q$-brane charge which are characterized by
the associated $q+1$-form conserved current. It is convenient to
dualize it to a closed $d-q-1$-form $J$.  We define a finite
``charge'' by integrating
 \eqn\chargedef{Q(\Sigma_{d-q-1})=\int_{\Sigma_{d-q-1}}J~~,}
where $\Sigma_{d-q-1}$ is a closed $d-q-1$ subspace.  Note that this
definition of the charge coincides with the standard definition for
$q=0$ where $\Sigma_{d-q-1}$ is all of space. Shifting the current
by an exact $d-q-1$-form does not affect its conservation and does
not change the charges \chargedef.  This is known as an improvement
transformation.} in the supersymmetry multiplet.  Such nontrivial
string currents imply the existence of cosmic strings. Hence our
general discussion of cosmic strings gives a new perspective on the
conclusions of \refs{\KomargodskiPC,\KomargodskiRB,\SeibergQD}\ and
extends them.

Throughout most of this paper, we will neglect the
gravitational back-reaction of cosmic strings.  In all the models we
consider, that back-reaction is a simple deficit angle much smaller
than $2\pi$, and this certainly appears justified.

\newsec{Comments on gauge symmetries, observables and dynamical branes
in quantum field theory}

One theme of our paper is the description of quantum field theory
and quantum gravity models using their long distance behavior.  In
this section we will make some general comments (many of them are
well known) about the long distance properties of gauge theories.

\subsec{Continuous Abelian gauge theories}

A simple question, which often causes confusion, that one can ask
about an Abelian gauge theory in continuum quantum field theory, is whether the gauge
group is $U(1)$ or $\R$. It has long been known to {\it cognoscenti}
that this question can be answered by specifying the list of allowed
observables in the model.  We can have Wilson loops
\eqn\Elec{W_A(\Sigma_1, n_A) = \exp\left(i n_A \oint_{\Sigma_1} A
\right) ~~.} In a $U(1)$ gauge theory only $n_A\in \Z$ are allowed,
while in an $\R$ gauge theory every $n_A\in \R$ is allowed.
Correspondingly, in a $U(1)$ gauge theory we can have 't Hooft
operators
 \eqn\Magl{T_A(\Sigma_1,
m_A) = \exp\left(i m_A \oint_{\Sigma_1} V\right) } where $V$ is the
dual photon and $m_A \in \Z$.  Alternatively, we can define $T_A(\Sigma_1, m_A)$ by
removing a tubular neighborhood of $\Sigma_1$ which is locally
$\R\times \S^2$ and impose $\int_{\S^2} F = 2\pi m_A$.

Equivalently, we can characterize the compactness of the gauge
group, by specifying the allowed fluxes of $F$ and $*F$ through
various cycles.  The allowed fluxes and Wilson/'t Hooft lines are
restricted by the Dirac-quantization condition, but in quantum field theory nothing
requires us to include {\it all} charges consistent with the Dirac
condition as dynamical objects in the theory.  Some of the Wilson
and/or 't Hooft lines may just be non-dynamical probes of the
theory.

The relation between dynamical particles and probes can be usefully
thought of as an infinite mass limit.  If we take the mass of some
charged field to infinity, we can find the effect on low energy
fields due to virtual particles of large mass by the usual Wilsonian
methods. However, one can analyze more general states, containing
one or more large mass particles, by using the particle path
expansion of the functional integral over the massive field
\SchwingerNM.  For the case of Abelian gauge theories in a
perturbative phase, the particle path expansion involves Wilson
loops for heavy electric charges and 't Hooft loops for magnetic
charges\foot{In perturbation theory, only mutually local dyons can
be light simultaneously, and by convention we call those electric
charges.}.

We end this subsection with a comment about non-compact gauge
groups. One simple way to guarantee that one is talking about a
non-compact gauge group is to insist that the model has two
relatively irrational charges, say $1$ and $\sqrt{2}$.  We would
like to point out that any gauge invariant Lagrangian for this pair
of charges also has a global Abelian symmetry under which just the
field of charge $\sqrt{2}$ rotates. If we accept that there are no
global conserved charges in quantum gravity, then such a model
cannot be coupled to gravity. In section $4$ we will see that global
charges and irrational charges are both ruled out by the same
argument about black hole physics.  Of course, in field theory with gauge group $\R$ we include Wilson loops of irrational charge in the list of
observables, even if all dynamical charged fields have only integer charges.
Alternatively we can insist that we do not allow magnetic flux when
we study the theory on $\S^2 \times \R^2$. In quantum gravity, we
will see that the {\it completeness hypothesis} does not allow us to
make such choices. $\R$ gauge theories would have to allow
irrational electric charges, and this would lead to a forbidden
global quantum number, as above. In other words, {\it $\R$ gauge
fields do not exist in the quantum theory of gravity}.  All
continuous gauge groups are compact.

\subsec{The low energy description of $\Z_p$ gauge theories and
an emergent $\Z_p$ one-form gauge symmetry}

The four dimensional $BF$-theory with the Lagrangian \HorowitzNG
\eqn\BFlag{{i p \over 2\pi} B \wedge dA } is known to be a simple
topological field theory which does not have any local degrees of
freedom.  Following reference \MaldacenaSS\ we now show that this
theory is equivalent to another topological field theory -- a $\Z_p$
gauge theory.

A simple way to establish the relation of \BFlag\ to a $\Z_p$ gauge
theory  \MaldacenaSS\ is to start with the Lagrangian
 \eqn\firlag{ t^2(d \phi - p A)\wedge *(d \phi - p A) + \CL(A),}
which describes the Higgsing of a $U(1)$ gauge theory with gauge field $A$
by a Higgs field $\phi$.  $\phi$ is subject to the identification $\phi\sim
\phi + 2\pi$ such that $\exp(i\phi)$ carries charge $p$ under the
gauge group and hence the $U(1)$ gauge symmetry is Higgsed to
$\Z_p$.  We are interested in the low energy limit of this theory
which is obtained in the limit $t \to \infty$.  In this limit $A=
{1\over p} d\phi$ and therefore the low energy theory does not
include local degrees of freedom.

In order to analyze the low energy $\Z_p$ gauge theory in more
detail we start with \firlag\ and dualize $\phi$ to derive the
Lagrangian
 \eqn\firlagd{ {1\over (4\pi)^2 t^2}  H\wedge * H +  {i
p\over 2\pi}  B \wedge dA + \CL(A)~~,}
where $H=dB$.\foot{If $\CL(A)=0$ we can shift
 \eqn\shiftA{A\to A + {i \over 8 \pi p t^2}*H}
and remove the
first term in \firlagd\ showing that the Lagrangian is only $ {i
p\over 2\pi} B \wedge dA$. However, in the presence of observables,
this shift does not eliminate the dependence on $t$.}

If \eqn\lofA{\CL(A) = {1 \over 2 e^2} F\wedge *F~~,} we can further
dualize $A$ to another one-form $V$ with the Lagrangian
\eqn\firlagdd{ {1\over (4\pi)^2 t^2}  H\wedge *H  + {e^2 \over 8
\pi^2} (dV - pB)\wedge *(dV - pB)~~.} In this form we interpret the
vector field $V$ as a matter field, which Higgses the $U(1)$ gauge
symmetry of $B$ down to $\Z_p$.

Now we can take the low energy limit ($t \to \infty$) and if
$\CL(A)$ depends only on $F=dA$ (i.e.\ there is no Chern-Simons term
or charged matter fields), we end up with the low energy Lagrangian
\eqn\BFlaga{{i p\over 2\pi}  B \wedge dA } which is a $BF$-theory
with coefficient $p$.

It is important that the gauge fields $A$ and $B$ have their
standard $U(1)$ gauge transformation laws with standard
periodicities.  In particular, for every closed two-surface
$\Sigma_2$ and for every closed three-volume $\Sigma_3$
\eqn\FHquan{\eqalign{ &\oint_{\Sigma_2} F \in 2 \pi \Z \cr
&\oint_{\Sigma_3} H \in 2 \pi \Z ~~.}} Note that the equations of
motion of \BFlag\ set $F=H=0$ and therefore, \FHquan\ might look
meaningless.  This is not correct. \FHquan\ can be interpreted as a
rule defining which bundles are included in the path integral, and
the allowed gauge transformations on these bundles. When the path
integral is used to compute expectation values of line and surface
operators, these rules have content because $F$ and $H$ have delta
function contributions at the location of the operators.

We conclude that the topological field theory Lagrangian \BFlag\ can
be interpreted as a universal description of a $\Z_p$ gauge theory.
In fact, it also shows that an ordinary $\Z_p$ gauge theory must be
accompanied by a $\Z_p$ one-form gauge symmetry (recall our conventions of labeling
higher form gauge symmetries by the dimension of the gauge parameter).  From the
perspective of our starting point \firlag\ this new one-form gauge
symmetry is an {\it emergent gauge symmetry.}  It arises, like most
emergent gauge symmetries, out of a duality transformation.  Note
that both gauge fields are $U(1)$ gauge fields, but as we will see
below, the distinct observables are labeled by $\Z_p$.

We point out that this observation is consistent with the result of \WittenYA\
that the theory with $p=1$ is trivial.

What are the observables in the theory \BFlag?  The local gauge
invariant operators are
\eqn\localBF{\eqalign{
&d\phi - pA \sim *H \cr
& dV - pB \sim *F ~~.}}
However, the equations of motion in the low energy theory \BFlag\
show that they vanish; i.e.\ their long distance correlation
functions are exponentially small in the parameter $t$ or other
relics of short distance physics.

We can also consider electric (Wilson) operators.  For every closed
line $\Sigma_1$ and for every closed two-surface $\Sigma_2$ we study
\eqn\Elecop{\eqalign{ &W_A(\Sigma_1, n_A) = \exp\left(i n_A
\oint_{\Sigma_1} A \right) \cr &W_B(\Sigma_2, n_B) = \exp\left(i n_B
\oint_{\Sigma_2} B \right) }} with integers $n_A$ and $n_B$.  They
are interpreted as the effect of a charge $n_A$ particle with
world-line $\Sigma_1$ and a string with charge $n_B$ with
world-sheet $\Sigma_2$.

It easily follows from the equations of motion of \BFlag\ that every
one of these operators induces a holonomy for the other gauge field
around it.  In particular \eqn\corrf{\Big\langle W_A(\Sigma_1,{n_A}
) W_B(\Sigma_2,{n_B} )\Big\rangle \sim \exp\left(2 \pi i{ n_A n_B
\#(\Sigma_1,\Sigma_2)\over p}\right) ~~,} where
$\#(\Sigma_1,\Sigma_2)$ is the linking number of $\Sigma_1$ and
$\Sigma_2$.  Therefore only $n_A\mod(p)$ and $n_B\mod(p)$ label
distinct operators.  Alternatively, the fact that only $n_A\mod(p)$
and $n_B\mod(p)$ are important follows from performing singular
gauge transformations.  A $U(1)$ gauge transformation on $A$, which
winds around $\Sigma_2$ (and is singular on $\Sigma_2$) shifts $n_B
\to n_B + p$.  Similarly, a $U(1)$ gauge transformation on $B$,
which winds around $\Sigma_1$ shifts $n_A \to n_A+p$.

In terms of the equivalent $\Z_p$ gauge theory the operator $W_A$ in
\Elecop\ is simply the $\Z_p$ gauge theory Wilson line; i.e.\ it
describes the world-line of a probe particle in the representation
labeled by $n_A\mod(p)$ of $\Z_p$.  The operator $W_B$ in \Elecop\
represents the world-sheet of a probe string which is characterized
by the $\Z_p$ holonomy around it being $\exp\left({2\pi i n_B \over p}\right)$.  Clearly, this interpretation is consistent with \corrf.

We can also attempt to construct magnetic ('t Hooft) operators.
Naively, the operator $\exp(i \phi(P))$ at a point $P$ is an ``'t
Hooft point operator'' for $B$ and $\exp\left( i \oint_{\Sigma_1} V
\right)$ is an 't Hooft line operator of $A$.  However, these
operators are not gauge invariant under the $U(1)$ gauge symmetries
of $A$ and $B$ respectively.  In order to make them gauge invariant,
we attach $p$ open line electric operators to $\exp(i \phi)$ and
consider $\exp\left(i\phi(P) + i\sum_i \int_{l_i} A\right)$, where
all the lines $l_i$ end at $P$, which is gauge invariant. Similarly,
we attach $p$ open surface operators to $\exp\left( i
\oint_{\Sigma_1} V \right)$ to make it gauge invariant.  Special
cases of these configurations are the two 't Hooft operators
\eqn\thooftAB{\eqalign{ &T_B(P,m_B) = \exp\left[im_B\left(\phi(P) +p
\int_l A \right)\right] \cr &T_A(\Sigma_1,m_A) =
\exp\left[im_A\left(\oint_{\Sigma_1} V +p \int_L B \right)\right]~~,
}} where the line $l$ ends at the point $P$ and the surface $L$ ends
at $\Sigma_1$.  The other ends of these open line and open surface
can be on another magnetic operator or can be taken to infinity.
Instead of using $\phi$ and $V$ in \thooftAB\ we could have cut out
of our space an $\S^3$ around $P$ and a tubular neighborhood around
$\Sigma_1$, which is locally $\S^2 \times \R$, and impose $\int
_{\S^3} H = 2\pi m_B$ and $\int _{\S^3} F = 2\pi m_A$.  This way the
operators \thooftAB\ are expressed only in terms of the variables
$A$ and $B$.

However, these constructions of the operators \thooftAB\ shows that
their long distance behavior is in fact trivial.  First, note that
the charge $p$ line and surface in \thooftAB\ are invisible -- they
are like Dirac strings.  Second, we can always locally gauge $\phi=
V=0$. Equivalently, as we remarked above the equations of motion of
the long distance theory \BFlag\ set $F=H=0$ and hence we cannot fix
any nontrivial integrals of them to be non-vanishing.

We conclude that the only significance of the magnetic operators
\thooftAB\ is that a number of electric operators can end at the
same point provided their electric charges sum up to zero modulo
$p$.  Other than that, the operators \thooftAB\ are trivial in the
low energy $\Z_p$ theory.

The Wilson loop operators may be viewed, in a familiar fashion, as
the remnants of high mass particles of charge $1\leq k < p$ in the
low energy theory.  For example, we could include in the original
Higgs model, from which we derived our $BF$-Lagrangian, fields
$\psi_k$ of charge $k$ with a mass of order the Higgs VEV or higher.
These particles can be represented using the
Schwinger particle path expansion.  If $k$ is a factor of $p$, then
${p\over k}$ of these paths can end at a point, indicating the
existence of the gauge invariant operator $\psi_k^{p\over k} e^{ - i
\phi}$ in the underlying theory at the scale of the VEV.  Similarly,
the fact that $p$ elementary flux tubes of the $BF$-theory can end
at a point and are therefore unstable (in the $BF$-theory, this just
means that their AB-phase is unobservable, so they are
indistinguishable from the vacuum), could be attributed to the
existence of magnetic monopoles carrying the minimal Dirac unit with
respect to the particles with charge $k = 1$.  Note however that in
an underlying continuum Higgs model, nothing requires these
monopoles to exist at finite mass.  In the $\Z_p$ lattice gauge
theory they do exist, as well as in continuum theories in which the
$U(1)$ is embedded in a simple spontaneously broken non-Abelian
gauge theory.  We will see in section 4 that the completeness
hypothesis of quantum gravity, guarantees the existence of these
particles below a mass of order Planck scale.

\subsec{Example 1: The 4d Abelian Higgs Model}

We have seen that the universal low energy description of $\Z_p$
gauge theories, involves a KR gauge field, and that the
observables include cosmic string sources. We now turn to two
examples of explicit microscopic models in which such strings are
present.

The first is the $U(1)$ gauge theory with a charge $p$ scalar.
The short distance Lagrangian is
\eqn\UVLag{\CL =
|(\partial - i p A)\Phi|^2 - V(|\Phi|) +{1 \over 2 e^2} F\wedge *F ~~.}
We take $V(|\Phi|)$ such that $\langle|\Phi|^2 \rangle = t^2 \not=0$.  This
vev Higgses the gauge symmetry to $\Z_p$ and the spectrum is gapped.
Writing $\Phi = t \exp(i \phi)$ in \UVLag\ we recover the Lagrangian \firlag.
Hence, the low energy description of this theory is given by the
universal Lagrangian \BFlag.

Given an explicit UV theory we can consider the operators
\Elecop.  First, the theory \UVLag, has only charge $p$
elementary quanta. Therefore, the nontrivial Wilson lines
$W_A(\Sigma_1,n)$ with $n \not= 0\mod(p)$ correspond to probe
particles rather than dynamical particles.  If, however, we add to
\UVLag\ additional massive particles with electric charge one, then
all these Wilson lines can be realized by dynamical particles.

It follows from \UVLag\ that the theory has smooth string
configurations.  The Higgs field $\Phi$ vanishes at the core of the
string, where the full $U(1)$ gauge symmetry is restored, and its
phase winds around the core. The low energy description of $k$
strings with world-sheet $\Sigma_2$ is given by the operator
$W_B(\Sigma_2,k)$. These strings can be detected by using the
electric Wilson operators $W_A(\Sigma_1,n)$ where $\Sigma_1$ circles
around the string world-sheet.

Next we consider the magnetic 't Hooft operators \thooftAB. Because
of the Higgsing by $\Phi$, magnetic monopoles are confined -- they
are attached to strings.  More precisely, since $\Phi$ has electric
charge $p$, the magnetic flux in our strings is $1/ p$ times the
fundamental unit of magnetic charge and a charge $m$ monopole is
connected to $mp$ strings.  This configuration is the one we
discussed above when we considered $p$ open surface operators.  We
see that $\langle T_A(\Sigma_1, m)\rangle$ exhibits an area law with
string tension $mp$ times that of the basic strings.  In accordance
with our general discussion above, this area law sets this operator
to zero in the long distance theory.

More interesting is the possibility of adding to our theory
dynamical massive magnetic charges.  If they carry the fundamental
unit of magnetic charge, the number of strings $k$ is not conserved
but $k\mod(p)$ is conserved.  This fact is consistent with our
general discussion above in which a low energy observer was
sensitive only to $k\mod(p)$ rather than to $k$.

\subsec{Example 2: Coupling a $\Z_p$ gauge theory to the $SO(3)$ $\sigma$-model
and topology tearing operators}

In our first example above the theory was gapped and the low energy
theory was only the $\Z_p$ gauge theory.  The universal behavior of
its correlation functions were determined by the low energy theory,
but additional details depended on the UV theory.  In particular,
some properties of the strings including their tension depended on
the details of the potential at $\Phi \approx 0$ where the $U(1)$
gauge symmetry was restored.  We now study a variant of this theory
in which the low energy $\Z_p$ theory is coupled to massless matter
fields.  Here, in addition to various possible massive excitations,
some charged particles and strings are constructed out of the low
energy matter theory.

We replace the theory \UVLag\ with a similar theory with two fields
$\Phi_{i=1,2}$ with charge $p$.  We can take the potential to
constrain
\eqn\potcon{|\Phi_1|^2 + |\Phi_2|^2 = t^2   ~~.}
This constraint guarantees that the $U(1)$ gauge symmetry is Higgsed
everywhere in field space to $\Z_p$.  This theory has been studied
in \refs{\PantevZS\PantevRH-\CaldararuTC,\SeibergQD} from various
points of view.

Naively, the low energy theory is simply the $SO(3)$ $\sigma$-model.
However, the unbroken $\Z_p$ gauge symmetry has interesting low
energy consequences.  In order to derive them we parameterize the
two fields $\Phi_{1,2}$ subject to the constraint \potcon\ as
\eqn\phionetwo{\eqalign{ &\Phi_1=t{z e^{i\phi}\over \sqrt{1+ |z|^2}}
\cr &\Phi_2=t{e^{i\phi}\over \sqrt{1+ |z|^2}} ~~.}} Here $z$
parameterizes the $\S^2$ target space and $\phi$ can be changed by a
$U(1)$ gauge transformation.  Next, we dualize $\phi$ as in
\firlag\firlagd\ and shift $A$ as in \shiftA\ to find that the
$\sigma$-model couples to $A$ and $B$ through \eqn\lowensot{{i \over
2\pi} B \wedge (p F -\omega)~~.} where  $\omega$ is the pull back of
the K\"ahler form of the target space to space-time normalized such
that if we wrap the target space once, $\oint \omega \in 2\pi $.

We conclude that the $\sigma$-model is coupled to the $\Z_p$
gauge theory in the $BF$ presentation.

The equation of motion of $B$ sets $pF=\omega$.  In fact, up to a
gauge transformation we can  write \eqn\Aaeom{pA = a} where $a$ is
the pull back of the K\"ahler connection to space-time satisfying
\eqn\omegada{\omega = da~~.} Therefore the integral over every
closed two-surface $\Sigma_2$ in our space time must satisfy
\SeibergQD \eqn\infop{\oint_{\Sigma_2} \omega =p \oint_{\Sigma_2} F
\in 2\pi p \Z ~~.}

The paper \SeibergQD\ studied two classes of theories with
nontrivial constraints on the ``instanton number'' $\int_{\Sigma_2}
\omega$. First, the theory was coupled to a $BF$-theory, where $B$ played the
role of a Lagrange multiplier implementing the constraint on
$\omega$. Second, the constrained theory was derived as a $\Z_p$
gauge theory by using charge $p$ fields.  Here we see that in fact
these two presentations are dual to each other.

To explore the low energy content of our model, we study its allowed
operators. These include both ``matter'' components constructed out
of the $\sigma$-model field $z$ and $\Z_p$ components constructed
out of $A$ and $B$.

Let us start with the matter operators. First, we have all the
obvious local $\sigma$-model operators, which are constructed out of
functions of $z$ and its derivatives. Another local operator at a
point $P$, $\CO_{Hopf}(P)$ is constructed by removing a neighborhood
of $P$ from our spacetime and constraining the $\sigma$-model
variables on its $\S^3$ boundary to have a nontrivial Hopf map. Such
an operator is interpreted as creating a nontrivial particle -- a
``Hopfion'', at the point $P$.  We can also construct line operators
$\exp\left(i n \oint_{\Sigma_1} a\right)$ for integer $n$. Finally,
we can attempt to construct a line operator that represents the
creation of $m$ strings.  Ordinarily, it is defined by removing a
tubular neighborhood of the line whose boundary is locally $\R
\times \S^2$ and imposing $\int_{\S^2} \omega = 2\pi m$. However,
because of the condition \infop\ this is possible only for $m$ which
is a multiple of $p$.  In other words, only operators that create
$p$ strings can be constructed out of the $\sigma$-model variables.

Next we construct additional operators out of the gauge sector $A$
and $B$. We start with the Wilson line
\eqn\elecsmA{W_A(\Sigma_1,n_A) = \exp\left( i n_A \oint_{\Sigma_1}
A\right) = \exp\left(i {n_A\over p} \oint_{\Sigma_1} a\right)~~~.}
For $n_A$ a multiple of $p$ this operator was mentioned above as a
``matter operator.''  The appearance of the fraction $n_A\over p$ in
\elecsmA\ is interesting. It shows that this is not a standard
$\sigma$-model operator.  If we think of it as a result of a massive
particle of charge $n_A$ the fraction means that the field of this
particle cannot be described as a section of an integer power of the
line bundle over our $\sigma$-model target space.  It appears like
``the $p$'th root of that line bundle.'' The interpretation of this fact is obvious. The
massive field is not a section on the target space, which is
pulled back to spacetime. Instead, it is a section of a bundle on
spacetime, whose transition functions depend on the $\sigma$-model
fields. Since the $\sigma$-model configurations are restricted by
\infop, the corresponding massive field is well defined.

The second gauge sector operator is the surface Wilson operator
$W_B(\Sigma_2, n_B)$.  In the presence of this operator the equation
of motion $pF-\omega$ has a delta function on $\Sigma_2$. Hence this
operator represents the world-sheet of a string with $n_B$ units of
flux.  These strings can be interpreted either as $\sigma$-model
strings or as strings constructed out of massive degrees of freedom.

For $n_B$ which is not a multiple of $p$ the world-sheet $\Sigma_2$
must be closed.  If however, $n_B$ is a multiple of $p$, these
strings can end as in the second equation in \thooftAB.  Their end
can be described by the string creation operator made out of the
$\sigma$-model fields that we discussed above.  We conclude that in
this case we can use the matter degrees of freedom to construct an
analog of the second operator in \thooftAB.  However, since the
equation of motion of $A$ still sets $H=0$ there is no analog of the
first operator in \thooftAB.

Now suppose we add dynamical particles with magnetic charge $k$.  They are characterized by the integral
$\int_{\S^2}\omega = 2\pi pk $ around their world line. As above,
such particles, are confined -- they are attached to strings. The
wrapping number of these strings is $pk$.  If such excitations are
present, then a string with wrapping number $m$ can decay to a lower
winding number while preserving $m\mod(pk)$.  In particular, if
particles with $k=1$ are present, then only $m\mod(p)$ is conserved.
It is important to note that since $\int_{\S^2}\omega = 2\pi pk $
around a magnetically charged particle, creating or annihilating
such a particle has the effect of ``tearing the topology.'' We will
refer to these operators and analogous ones below as ``topology
tearing operators.''

As in the Abelian Higgs model, the expectation value of a purely
electric loop $\langle W_A(\Sigma_1,n)\rangle $ has a perimeter law.
If $\Sigma_1$ is large and it winds once around a string with
winding number $m$, $\langle W_A(\Sigma_1,n)\rangle$ depends on $m$
through a phase $\exp({2\pi i nm / p})$.  Again, only $m\mod (p)$ is
measurable this way.  Also, as in the Abelian Higgs model, if there
are no magnetically charged particles, an 't Hooft loop of charge
$m$, $T_A(\Sigma_1, m)$, exhibits an area law associated with a
string tension proportional to $ mp$. However, dynamical
magnetically charged particles with charge $k$ can screen the
tension and the area law is determined by $m \mod (k)$.  In
particular, for $k=1$ the loop can be totally screened.

In conclusion, we have presented new operators in the standard
$SO(3)$ $\sigma$-model.  In addition, this model can be coupled to a
topological field theory.  There are no new local degrees of freedom
but the set of operators is modified.  We have also shown that this
theory can be modified at short distance (without breaking its
$SO(3)$ global symmetry) by adding electric and magnetic charges.
The electrically charged fields do not transform like sections of a
line bundle on the target space.  The magnetic charges violate
conservation of the string current $*\omega$. Equivalently, because
of the monopoles the $2$-form operator $\omega$ is not closed
 \eqn\domega{d \omega= p dF \not=0.}
Hence, the topological conservation law is violated. As we said
above, this arises from the fact that creation of magnetically
charged particles tears the topology.

The topological classification of the space of maps from
space-time to the target space depends on the topology of both domain
and range, and violation of the conservation law can come from
either. For example, we can, as above, embed the $\sigma$-model in a
linear model in which the constraint \potcon\ is a low energy
artifact. This eliminates the topology of the target space, while
keeping space-time continuous.  Alternatively, we can look at the
lattice version of the $\sigma$-model, with the constraint imposed at
the microscopic level. Here, topology tearing operators and topology tearing excitations can ``hide
their singularities in the holes between lattice points'' and they
exist even though the target space topology is intact.  We will see
that in theories with gravity, black holes act in some respect like
both of these mechanisms at once.

Finally, we would like to comment on another aspect of this theory,
which is unrelated to our $\Z_p$ discussion. The $\sigma$-model has
particle like excitations, ``Hopfions'' associated with nontrivial
Hopf maps of space to the target space. Their number is
conserved\foot{We point out that the corresponding conserved current
is \eqn\Hopfcu{J_H \sim \sum_{i,j} \left(\bar \Phi^i d \Phi_i -
\Phi_i d \bar\Phi^i\right)d\bar\Phi^j d\Phi_j \sim d\phi \wedge
\omega~~,} where we have used the parametrization \phionetwo.  It is
not gauge invariant under the gauge symmetry of $A$ and therefore
this current cannot be gauged. (Equivalently, if we express it in
terms of the gauge invariant data of the $\sigma$-model as $a
\wedge\omega$, it is not invariant under the global $SU(2)$.)
However, the corresponding charge is gauge invariant and therefore
the Hopf number is conserved.} and hence the theory has a {\it
global continuous symmetry}.  This symmetry cannot survive when the
model is coupled to gravity. The operator $\CO_{Hopf}$ , when added
to the Lagrangian, will describe the breaking of this symmetry.
Although formally it is a local operator, we are dealing with a
non-renormalizable theory, so it is at best local on the scale of
the cutoff $t$.  So we should imagine that the hole which it cuts in
space-time is of size $\sim 1/ t$.  A smooth Hopfion configuration
can shrink to the cutoff scale and when, in first order perturbation
theory in $\CO_{Hopf}$, it encounters the hole, it disappears. Its
energy is radiated into topologically trivial fields of the
$\sigma$-model.  If in fact the underlying physics responsible for
violation of Hopf number comes from a shorter distance scale, the
dimensionless coefficient of $\CO_{Hopf}$ in the Lagrangian will be
small.

There is a similar
description of proton decay, due to {\it e.g.}\ unification scale
exchanges, in terms of a topology tearing operator in the Skyrme
model.  We will discuss it in Appendix B.

\newsec{The taxonomy of stable strings}

In this section we review and expand on Polchinski's classification
of cosmic strings \PolchinskiBG.  Polchinski classifies strings
according to a long distance observer's ability to detect them:
\item{1.} ``Local'' strings cannot be detected by a long distance observer.
\item{2.} ``Global'' strings are coupled to a massless $2$-form gauge field $B$ or equivalently to a massless dual scalar $\phi$, which winds around the string.
\item{3.} Aharonov-Bohm (AB) strings \refs{\AlfordSJ\KraussZC-\PreskillBM}
are characterized by a nontrivial discrete holonomy around the
string.  We can divide this category in two by asking whether or not
the discrete group is Abelian. We will leave a discussion of the IR
theory for discrete non-Abelian groups to future work.
\item{4.} Quasi-Aharonov-Bohm strings have such a holonomy but it
is embedded in a continuous gauge symmetry.

We would like to make several comments about these classes, emphasizing
the perspective of this paper:
\item{1.} The term ``local strings'' originates from the cosmic string
literature.  From our perspective this terminology is confusing
because such strings are characterized by a {\it global string}
charge, which is not gauged.  If the corresponding symmetry is
continuous, the theory has a conserved string current -- a closed
but not exact $2$-form current $J$.  Following the three quantum
gravity conjectures, such global string symmetries cannot be present
and, as discussed in \PolchinskiBG, such stable strings should not
exist in a gravity theory.
\item{2.} Again, the term ``global strings'', which
originates from the cosmic string literature is confusing.  Such
strings are characterized by a {\it local string} charge.  We will
discuss them in more detail below.
\item{3.} Here the Aharonov-Bohm (AB) phase is associated
with a {\it discrete gauge symmetry}, e.g.\ $\Z_p$.  The discussion
of section 2 shows that these are $\Z_p$ strings.  They arise from
an ordinary $\Z_p$ gauge symmetry and are coupled to an emergent
$\Z_p$ one-form gauge symmetry.  So the difference between this case
and the previous one is that here the one-form gauge symmetry is
discrete.  As we will discuss in section 4, in the context of a
gravity theory $p$ such strings can combine and disappear.
\item{4.} Unlike the previous case, here the ordinary gauge symmetry
is embedded in a continuous gauge symmetry, e.g.\ $\Z_p \subset
U(1)$.  Since the low energy theory does not have a $\Z_p$ gauge
symmetry, we do not find the emergent $\Z_p$ one-form gauge symmetry
and hence these strings are at best associated with a global charge
and should not be present in quantum gravity.  In fact, in most
situations such strings can decay by emitting the massless $U(1)$
gauge fields.

\bigskip
To summarize, classes 2 and 3 are associated with a long range
one-form gauge symmetry, which is continuous and discrete
respectively, while cases 1 and 4 are not associated with such a
gauge symmetry and hence cannot exist in a consistent model of
quantum gravity. We have given a complete IR description of class 3
strings when the discrete AB-phase is Abelian. The non-Abelian case
certainly exists, because we know examples with discrete non-Abelian
gauge groups in perturbative string compactification (e.g.\ the ${\bf S}_5$
symmetry of the quintic is a case in point). We will leave a
discussion of the IR Lagrangian for such theories to future work.
For the rest of this paper we restrict the AB-phase to lie in $\Z_p$.

We now turn to a more detailed discussion of the second class in
which the strings are coupled to a massless scalar $\phi$, which
winds around the string.  This massless $\phi$ leads to two
problems:
\item{1.} As emphasized in \PolchinskiBG, such a configuration
of a winding $\phi$ is possible only if $\phi$ does not have a
potential.  Typically, this is natural only if the theory has a
global shift symmetry of $\phi$.  The lack of global symmetries in
gravity makes this possibility unnatural.
\item{2.} Even if a scalar $\phi$ without a potential is present,
its kinetic term $(\partial \phi)^2$ makes the tension of the string
infinite.

\bigskip
These two problems can be solved naturally, if our theory and our
vacuum are supersymmetric.  In that case it is natural to have no
potential for $\phi$ without a shift symmetry. $\phi$ and its
supersymmetric scalar partner lie on a supersymmetric moduli space.

In the context of supersymmetry the problem of infinite tension can
also be ameliorated. $\phi$ is typically the real part of a massless
complex field $\tau$ without a potential.  More generally, we should
view $\tau$ as a complex coordinate on the moduli space. For
simplicity, consider a common example of a one dimensional
moduli space where the metric on that space is determined by the
kinetic term \eqn\taukinet{g_{\tau\bar \tau} \partial\tau \partial
\bar \tau = {1\over ({\rm Im} \tau)^2} \partial \tau \partial
\bar \tau~~.} Because of the singularity as ${\rm Im} \tau \to
\infty$, if ${\rm Im} \tau \to \infty$ near the boundary of
space-time, the string tension can be finite.\foot{Using polar coordinates in two dimensions, the most general solution of the equations of motion, which winds around infinity $n$ times is $\tau= - {n\phi \over 2\pi}+ i {n\over 4\pi C} \left((br)^C - (br)^{-C}\right)$, where $b$ and $C$ are integration constants.  The action of this solution suffers from an IR divergence as $r\to \infty$.  It is removed by taking the $C\to 0$ limit in which the solution becomes $\tau = i{n\over 2\pi } \log (bz)$ where $z=re^{i\phi}$.  Note that this solution becomes singular at $z=1/b$.  In standard string applications this singularity is is removed by a duality transformations.  We thank J.~Distler for a useful discussion about this point.}

One can generalize this construction to multi-dimensional moduli
spaces.  Finite tension strings will exist for every non-trivial
circle {\it near the boundary of moduli space}, if the metric has
the appropriate asymptotic behavior there. To our knowledge, this is
the case for all the boundaries of moduli space that have been
explored in string theory.  Note that it does not matter whether the
circle can be unwound in the interior of moduli space. The existence
of a string is characterized by the winding of $\phi$ at spatial
infinity, but the finite tension condition forces the fields to the
boundary of moduli space at spatial infinity. This is analogous to
the finite action condition on four dimensional gauge theory, which
leads us to characterize gauge fields in terms of maps of the sphere
at infinity into the gauge group.

This understanding leads us to an important point: the behavior of
these finite tension strings is such that they cannot be viewed as
excitations of a model at a generic point in moduli space. Generic
models are characterized by generic values of the moduli at infinity
and such boundary conditions do not accommodate finite tension
strings.

At this point we would like to make a few comments:
\item{1.} For generic values of the moduli we can consider
a circular loop of one of these strings, of radius $R$.  Far from
the string the moduli must first approach the boundary of moduli
space, and then return to their fixed value at infinity.   As $R
\rightarrow \infty$, the fields must return to the fixed value at
infinity over a larger region in space and hence the energy of this
configuration must grow more rapidly than $R$.  (Of course, this
statement is equivalent to our statement above that such strings
have infinite tension unless the asymptotic value of the moduli is
infinite.) Consequently, for large enough $R$ this configuration
collapses into a black hole, and there are no large loops of cosmic
string.  For example, in the simple moduli space with asymptotic
metric \taukinet, the energy grows like $R {\rm ln} R$ and the
maximal scale of the string loop is of order $\mu^{-1} e^{ {m_P^2}\over
{\mu^2}}$, where $\mu^2 $ is the string tension.
\item{2.}  Special cases of this discussion of cosmic strings
with a $\Z$ valued quantum number in four asymptotically flat
dimensions are fundamental Type II or heterotic $E_8 \times E_8$
superstrings, in compactifications that preserve minimal
super-Poincare invariance in four dimensions.  The infinite
tension in these cases is of order $g_s^2$ as is the back reaction
on the metric.  For this reason this effect is often ignored.
\item{3.} We would like to point out an important fact,
which follows from the analysis of \KomargodskiPC, and to which we
will return in section 5.  The target space of every supersymmetric
field theory is a K\"ahler manifold.  If its K\"ahler form is not
exact, it cannot be coupled to linearized minimal supergravity.  The
only way to do that is through the introduction of an additional
chiral superfield $\rho$.  In the linearized approximation the
K\"ahler potential $K$ is replaced by $K+\rho+\bar \rho$.  The real
field ${\rm  Im}\rho $ plays the role of $\phi$ above -- it is dual
to a $2$-form that couples to the string current.  However, taking
into account the gravitational corrections to the metric on the
target space, this modification of $K$ has the effect of ruining the
topology of the target space. The strings are no longer stable, even
classically.
\item{4.} At this point it is worth mentioning the
gravitational back-reaction of the cosmic string, which is a deficit
angle at infinity, and has been neglected in all of our discussions
so far.  This by itself means that there is a sense in which the
string is not an excitation in asymptotically flat space, but if the
deficit angle is small this is clearly sort of academic.  One can
make arbitrarily large closed loops of such strings without forming
black holes.  In particular, if the tension of $\Z_p$ strings is
small compared to the Planck scale, then macroscopic strings are
observable.
\item{5.} We caution again that the phenomenological import of our statements
is unclear.  Many of the instabilities we describe may have
exponentially small probabilities.  For example, our general
arguments only imply the existence of magnetic monopoles of mass
$\sim M_P$.   In that case the Schwinger pair production probability
per unit string length for these monopoles to destabilize an
infinite cosmic string is  $\exp \left(- {M_P^2\over f^2}\right )$,
where $f$ is of the order of the energy scale of the magnetic flux
line in some low energy Higgs field. Similarly, the collapse of a
long but finite string for generic values of the moduli to a black
hole takes place only for exponentially large $R$. Therefore, in
many cases, these instabilities will have no phenomenological
relevance.

\bigskip
Our conclusion then is that in models of quantum gravity in four
asymptotically flat dimensions, with finite values of the moduli at
infinity, there is only one kind of stable cosmic string, and it is
stable modulo some integer $p$.\foot{There is a
generalization to strings whose AB-phases lie in a more complicated
finite group.} In models that have a non-compact moduli space, the
existence of strings of finite tension depends on the behavior of
the metric on target space at infinity and on the ``fundamental group of the
boundary.''  All other types of cosmic strings suffer from
instabilities. In the next section we will provide arguments for the
claims about quantum gravity, which we have used to come to these
conclusions.

\newsec{Symmetries in quantum gravity}

In this section we will discuss the following three conjectures
about symmetries in quantum gravity:
\item{1.} No global symmetries -- all the symmetries including
brane symmetries and discrete symmetries are gauged.
\item{2.} All continuous gauge symmetries are compact.
\item{3.} The entire set of allowed states in the lattice of charges is populated.

\bigskip
Before getting into a detailed discussion of these conjectures, we
would like to point out that some special cases of them can be
proven or at least be argued for quite convincingly, but others are
more speculative.  Furthermore, these conjectures are not logically
independent.  Assuming a subset of them, or some special cases, we
can prove the others.

As an example of the interrelation between the conjectures, consider
a gauge symmetry associated with a $q$-form conserved current.  Then
the corresponding field strength is a $d-q+1$ form $F$.  If the
gauge symmetry is noncompact, it does not have magnetic charges and
$F$ is closed and leads to a conserved current.  Using the first
conjecture, it must be gauged, i.e.\ be coupled to a $q-1$ gauge
field $B$ through the Chern-Simons coupling $B\wedge F$.  As in
section 2, this has the effect of Higgsing the two gauge symmetries.
If such a coupling is not present, we conclude that the gauge group
must be compact and that the theory includes dynamical magnetic
monopoles, so that $F$ is not closed.  A similar example was
discussed at the end of section 2.1.  There we argued that if a
gauge group is $\R$ and two relatively irrational charges are
present, the theory must have a global symmetry, thus violating the
first conjecture.

\subsec{No global continuous symmetries}

The idea that any theory of quantum gravity cannot have global
symmetries has a long history and is often referred to as a
``folk-theorem.''  Below we will discuss the rationale behind this
idea.  But before doing that we would like to make some comments:
\item{1.} The simplest form of this conjecture applies to continuous symmetries,
which are associated with point particles in space-time dimensions
four or larger.  The usual argument for this case involves black
hole physics.
\item{2.} The lack of such ordinary global continuous symmetries
is known to be satisfied in all controlled constructions of quantum
gravity.  It was shown in \BanksYZ\ to be satisfied in perturbative
string theory -- global symmetries on the string world-sheet lead to
gauge symmetries in spacetime, and there is no way to have global
symmetries in spacetime.  The situation for strings in AdS
background \AharonyTI\ is similar -- global symmetries in the
boundary theory are associated with gauge symmetries in the bulk and
there is no way to have global symmetries in the bulk.  Similar
comments apply to the matrix model of \BanksVH.
\item{3.} Assuming that no such global symmetries exist, we can
easily extend the result to all continuous symmetries associated
with branes of codimension larger than two.  Such branes can be
compactified on circles to ordinary point particles in lower
dimensions.  The absence of global symmetries for these particles
leads to the absence of global symmetries of the higher dimensional
branes.
\item{4.} We will be particularly interested in string currents in four
dimensions.  These situations are not covered by the discussion
above.  Yet, we will argue in section 4.4 that such currents should
also be coupled to gauge fields.  One special string current will be
discussed in section 5.  This current appears in the supersymmetry
current algebra.  Hence, in the context of supergravity it must be
gauged, independent of our more general conjectures.

\bigskip
We begin by reviewing the argument that models of quantum gravity
cannot have global continuous symmetries. The lightest particle
transforming under the group, is in a representation ${\bf r}$ and has mass $m$.  We can make multi-particle states of an arbitrary representation ${\bf R} \subset {\bf r}\otimes {\bf r}\otimes {\bf r}...$ and
collide those particles to make a black hole. As long as the
representation is not correlated with a long range gauge field, this
will be a Schwarzschild black hole, and its macroscopic fields will
be independent of ${\bf R}$.  Hawking's calculation shows that this state
decays to a remnant with mass of order $ X M_P$, without emitting
its global charge. Here $X L_P$ is the value of the Schwarzschild
radius for which Hawking's calculation becomes unreliable, . The
value of $X$ for which the Hawking calculation breaks down is hard
to estimate but it is surely no more than an order of magnitude or
so.

If $m$ is non-zero, the remnant state is absolutely stable, since any
combination of particles carrying such a large representation of the
global group is heavier than the remnant. Even if $m = 0$, as would be
the case for the global $SU(8)$ symmetry of $N=8$ SUGRA, the
lifetime of the state must go to infinity with the number of particles in ${\bf r}$, which are needed to make a black hole in ${\bf R}$. Indeed, the only
way to emit an infinite number of massless particles with finite
energy is through bremsstrahlung processes in which the decay products accelerate through
re-scattering. Neither the remnant state, nor any of its decay
products carry charge under any low energy gauge field apart from
gravity itself. Gravitons are
of course neutral under the global symmetry, so gravitational
bremsstrahlung cannot carry away global charge. Thus, even if $m=0$
we get an infinite number of arbitrarily long lived remnants, whose
external geometry is that of Schwarzschild with $R_S = X L_P$.

The Covariant Entropy Bound (CEB) \refs{\FischlerAB,\BoussoXY} is a
conjectured bound on the entropy contained in any causal diamond in
Lorentzian geometry. The original statement of the bound does not
specify which density matrix is (implicitly) referred to, but it
would be completely meaningless if it did not count entropy
associated with verifiably different states that could form the same
black hole geometry. Causal diamonds whose past boundary is a
portion of the remnant black hole horizon\foot{The black hole
remnant might be only meta-stable in the massless case and purists
might object to using its interior geometry. We can instead talk
about the causal diamond of an observer orbiting the black hole in
the last stable orbit, over one period of rotation.  This would
affect only the quantitative estimate of the bound on the order of
finite global symmetry groups, which we present below.} have a
holographic screen whose area is that of the horizon.  Thus, the CEB
implies that the entropy of the remnant object is less than a finite
number of order $\pi X^2 $, which contradicts the existence of an
infinite number of remnant states. Other arguments have been given
for the absence of infinitely degenerate remnants, but all of them
make {\it some} assumption about the most general possible theory of
quantum gravity. For example, it has been argued that quantum loops
of an infinite number of degenerate remnants are inconsistent. All
such arguments contain {\it some} hidden assumptions. We view the
CEB as the most elegant and general criterion for characterizing the
differences between gravitational and non-gravitational quantum
theories.

The CEB also enables us to get a soft bound on the maximal size of a
{\it finite} global symmetry group. Let $\Sigma $ be the sum of the
dimensions of irreducible unitary representations of such a group.
Since each state, in each representation, would be another possible
state for the remnant, we get a bound \eqn\oGXs{\Sigma < e^{\pi X^2}
~~.} The bound is soft, both because we do not know the value of
$X$, and because for $X$ of order one we expect corrections to the
Bekenstein-Hawking formula.

Actually, as we said above and as we will discuss in section 4.2, it
is our prejudice/conjecture that there are no global discrete
symmetries in gravitational theories.

We want to point out that this set of arguments also rules out
Abelian gauge fields with particles of relatively irrational charge.
Suppose for example that there are particles of charge $1$ and
$\sqrt{2}$. Then we can make Reissner-Nordstrom black holes of
charge \eqn\qirr{q = n_1 + \sqrt{2} n_2 ~~,} where the $n_i$ are any
integers. Thus, there are an infinite number of black holes of
charge $q < \epsilon$, for any $\epsilon$.  Again, Hawking radiation
will allow all of these to decay down to a remnant of Schwarzschild
radius $X L_P$, but an infinite number of these states are
indistinguishable, violating the CEB.  This argument complements
that given in section 2.1, that the theory has a global $U(1)$
symmetry, which counts the number of particles of charge $\sqrt 2$.
The argument for compactness of continuous symmetry groups and the
fact that they must be gauged, are really one and the same.

\subsec{Are discrete symmetries gauged?}

In perturbative string theory there is an enormous amount of
evidence that all discrete symmetries are gauge symmetries.  For
example, the $SL(2,\Z)$ duality symmetry of ten dimensional Type IIB
string theory is, from the M-theory point of view, just part of the eleven-dimensional reparametrization group.  Similarly, the T-duality of
Heterotic strings compactified on a circle, is a gauge symmetry
\DineVU, because it is part of a continuous $SU(2)$ gauge symmetry
that is restored at the self dual radius.  Some of the torsion part of the
K-theory group of D-brane systems can be viewed as part of the gauge
group of a space-filling brane anti-brane system out of which we
construct lower dimensional $D$-branes \refs{\MinasianMM\SenRG\SenII\SenSM-\WittenCD}. For example, the
$\Z_2$ of the stable non-BPS particle of Type I string theory is the
center of the $Spin(32)/\Z_2$ gauge group of the branes. And so it goes
for all known discrete symmetries.

We anticipate a general argument for the fact that discrete
symmetries are gauged, based solely on simple principles of any
theory of quantum gravity, but we have not found one as compelling
as those for continuous symmetries.  The best we could do is to find
the argument around \oGXs, which leads to a weak bound on the size
of possible discrete groups.

A first step in this direction might be to
establish that a $\Z_p$ symmetry is gauged if there is a finite
tension cosmic string, around which particles pick up a $\Z_p$
phase.  The existence of such a string would enable us to measure
the $\Z_p$ charge of a black hole, proving that the $\Z_p$ charge is
conserved and measurable from arbitrarily far away.
There is clearly much more to be understood about this.

\subsec{The Completeness hypothesis and black holes}

On the basis of perturbative string theory examples, Polchinski has
conjectured that consistent models of quantum gravity always contain
a complete set of electric and magnetic objects consistent with the
Dirac quantization condition \PolchinskiBQ. This is certainly not the case in
ordinary quantum field theory. There we can often take the masses of
monopoles or charged particles to infinity, leaving behind an
incomplete but consistent spectrum.

In a theory of gravity, this procedure will fail. When the mass of
the particle becomes larger than the Planck mass, gravitational
back-reaction cannot be neglected and the state becomes a black hole
carrying the electric or magnetic charge of the erstwhile particle.
This black hole will decay by Hawking radiation down to the extremal
black hole for that charge. Thus, there is no way to tune parameters
to eliminate charges.

Still, we might speculate about the possibility of models in which
certain allowed states were ``never there.''  In quantum field
theory, there are two ways of specifying the global structure of a
gauge group.  We can either specify the allowed set of observable including
line and surface operators, or we can specify the
allowed fluxes of the field strength and its dual on geometries
containing a non-trivial cycles.  The line operators require
renormalization by a factor
\eqn\ecing{e^{ c \int \sqrt{-g} }~~,}
and therefore interact with the gravitational field like a point
particle. Thus, an allowed line operator will produce a charged
black hole with the ``missing'' charge, even if the field theorist
has refused to include a low energy field carrying that charge in
the model.

Furthermore, the black hole geometry has an $\S^2$, which is
topologically non-trivial, and so the field theorist will be forced
to allow black holes with every flux that went into the definition
of the gauge group.

Without a rigorous definition of the most general possible theory of
quantum gravity, we cannot call these remarks a proof, but they
constitute a very strong argument that the completeness hypothesis
is valid in any such theory. As a consequence of the completeness
hypothesis, our argument about relatively irrational charges becomes
the stronger statement that {\it all continuous gauge groups are
compact}. The completeness hypothesis eliminates the possibility of
a non-compact gauge group (no monopoles), which ``just happens" to
have only quantized electric charge states.

Our argument that gauge groups must be compact applies only to
continuous groups. We know that Type IIB string theory has an
infinite $SL(2,\Z)$ gauge symmetry. We also have not found an
argument that bounds the value of $p$ in a discrete $\Z_p$ gauge
theory. Thus, we cannot rule out the $p \rightarrow \infty$ limit.
We note however that $SL(2,\Z)$ is always Higgsed to a finite
subgroup by a choice of the moduli fields at infinity.  Furthermore,
known string compactifications to asymptotically flat space always
seem to have finite discrete gauge groups, acting on the Hilbert
space of excitations. Thus, it may be that more refined arguments
than those we have presented could rule out infinite discrete gauge
groups that leave the boundary conditions at infinity invariant.

It is quite likely that eventually the completeness hypothesis would be derived from general string theory considerations without relying on the existence of black holes.  In fact, there are known simple solvable examples of string theory in one space or in one space and one time dimensions in which this question can be addressed.  Clearly, in such degenerate situations there are no black hole solutions.  So one could ask whether all allowed charges are present.  The relevant charges are the RR charges of the type 0 theory.  The authors of \refs{\DouglasUP\KlebanovWG\SeibergNM\SeibergEI-\MaldacenaHE} have shown that all the RR charges of these theories have a manifestation in the matrix models.  Furthermore, the allowed charges are quantized and {\it all} of them are populated.

\subsec{Black holes and the stability of strings}

In this subsection we focus on strings and follow the point of view of \PolchinskiBG, which emphasizes measurements, which can be performed far from the string.  We restate it as follows from the three conjectures: every stable string must be coupled to a KR gauge field.  Its gauge group can be either $U(1)$ or $\Z_p$.\foot{In this discussion we ignore the interesting situation which arises when strings are associated with a non-Abelian AB-phase.}  In the first case the total number of strings is conserved while in the latter case it is conserved only modulo $p$.  Let us see how these ideas are realized in two examples of strings: Nielsen-Olesen strings of the Abelian Higgs model (section 2.3) and $\sigma$-model strings (section 2.4).

The completeness hypothesis shows us that in quantum gravity
Nielsen-Olesen strings of low energy Higgs models are at best
unbreakable modulo $p$, for some integer $p$.  The value of $p$ is
determined by the charge of the Higgs field which in turn determines
the unbroken gauge symmetry.  The string can be detected by the
AB-phase around it.  The completeness assumption guarantees that the
system always has dynamical particles of electric charge one, which
can detect the $\Z_p$ phase.  Furthermore, while in field theory the
system might not have magnetic monopoles of the lowest allowed
charge, the completeness hypothesis now tells us that in gravity
theories such monopoles must be present.  $p$ strings can end on
such a monopole. Hence, not only can't we detect $p$ strings using
an AB-phase, but such a configuration of $p$ strings is unstable. In
terms of the $U(1) $ field strength we have \eqn\Fnonco{d  F =
J_m\not = 0 ~~,} where $J_m$ is the monopole current, so the local
string current is not conserved.

As we have explained in sections 2 and 3, one way to think about such a
situation is in terms of a continuum $\Z_p$ gauge theory, which is a
pair consisting of a $2$-form gauge field $B$ and a one-form $A$ which are both valued in $U(1)$.
The action is
\eqn\BFsecf{{ip \over 2\pi} \int\ B \wedge F ~~.}

Next we turn to $\sigma$-model strings.  One way to generate these $\sigma$-models is by using a Higgs system and then the strings are similar to those we have just discussed.  Alternatively, we can use a description which is intrinsic to the $\sigma$-model variables.  The strings are maps of a plane or a two-cycle in space into a non-trivial two-cycle in the target space.  Now consider a black hole space-time.  The black hole horizon is a two sphere and we can consider maps which take this two sphere into the
non-trivial two cycle in the target space. These can end a string
configuration without any violation of continuity.  Colloquially, we
can say that ``the other end of the string has fallen down the black
hole.''  We can make a particularly compelling case for the existence
of such string ends if the model contains a $U(1)$ gauge field. We
can study the space-time of an extremal charged black hole, which
has an infinitely long throat, ending in an $AdS_2$ horizon. It
would appear to violate locality to claim that one could not thread
a $\sigma$-model string through the throat.  If we now drop
charged particles into the extremal hole, we are left with the
Schwarzschild string ender, when the dust of Hawking radiation has
cleared.

As we explained in section 2.4, if the $\sigma$-model is coupled to a $\Z_p$ gauge theory through \lowensot, then there can be massive particles which are  not sections of a line bundle over the target space -- they appear like ``fractional powers of that line bundle.''  The completeness assumption guarantees that such dynamical particles exist.  Such particles can detect the $\Z_p$ AB-phase around the
string.  If such a string can end, this phase would change discontinuously as we pass the string end. Thus, string ends can exist only if they are not detectable by this phase.  In fact, the completeness assumption guarantees that such string ends must be present -- each is an end of $p$ strings.  Again, we can say: ``the number of strings, which can penetrate the black hole horizon is a multiple of $p$.''

We have seen that by coupling the $\sigma$-model to a $\Z_p$ gauge
theory, we can arrange for $\Z_p$ AB-phases in the interaction of
probe particles with the $\sigma$-model strings.  We conclude that
in a model of quantum gravity, these strings, like those of the
Higgs model, have no integer valued quantum number, but might be AB-strings for some finite group.

The violation of string current conservation by string ends is
analogous both to its violation in lattice $\sigma$-models, where
topology change can happen in the holes between lattice points, and
in models where the target space topology is a low energy artifact.
{}From the external observer's point of view, the black hole horizon
is a hole in space, on which we are free to specify boundary
conditions with non-trivial topology.  From the internal point of
view, the Hamiltonian is time dependent and singular, and all low
energy field theoretic restrictions must be abandoned.  The same is
true for an external observer supported very close to the horizon,
which experiences Hawking radiation of extremely high temperature.

Finally, we should add a word of caution.  Throughout this subsection we have ignored the back reaction of the gravitational field.  An infinitely long string in four dimensions creates a deficit angle at infinity and as such some of the analysis above is not necessarily justified.  If the string tension is parametrically smaller than the Planck scale, it is reasonable to ignore this back reaction.  It is not clear to us how to think about it for strings with Planck scale tensions.

\newsec{Relation to Supergravity}

So far supergravity has not played a role in our discussion. As we
will now see, the discussion above allows us to streamline and to
reinterpret some recent results about constraints on supergravity
\refs{\KomargodskiPC,\KomargodskiRB,\SeibergQD} and to further
extend them.  Throughout this section we will limit ourselves to
$\CN=1$ supersymmetry in four dimensions.  The extension to other
cases is straightforward.

An important distinction in
\refs{\KomargodskiPC,\KomargodskiRB,\SeibergQD} is between
supersymmetric theories containing a continuous parameter, which can
be dialed to make gravity arbitrarily weakly coupled, and theories
which are intrinsically gravitational, in the sense that all scales
are discrete multiples of the Planck scale. We will devote the next
two subsections to these two cases and will find that all of the problems
exposed in \refs{\KomargodskiPC,\KomargodskiRB,\SeibergQD} can be
phrased in terms of the existence of string currents. When
all such currents are gauged or their conservation laws violated,
following the strictures of section 4, we find rather different
outcomes in the two classes of models. In models with continuous
parameters, the duals of the KR gauge fields are new chiral
multiplets. These make all FI-terms field dependent, and remove all
non-contractible two cycles from the low energy target space. In
models with quantized FI-terms or periods, the KR gauge symmetry is
discrete, and the low energy theory includes our universal $\Z_p$
Lagrangian.

\subsec{Linearized SUGRA}

Following \KomargodskiRB\ we start with constraints on linearized
supergravity. Here one starts with a well defined rigid
supersymmetric theory and studies its  supersymmetry current
multiplet.  Then, this current is coupled to gauge fields -- the
supergravity multiplet.  This is a standard and well known procedure
(see e.g.\ the text books \refs{\GatesNR,\WeinbergCR}).  Depending
on the super-current multiplet three different approaches can be
taken:
\item{1.} Old minimal supergravity \refs{\StelleYE\FerraraEM-\FradkinJQ}
uses the Ferrara-Zumino (FZ) multiplet \FerraraPZ
    \eqn\FZmu{\eqalign{
    &\bar D^\alphadot \CJ_{\alpha\alphadot} = D_\alpha X \cr
    &\bar D_\alphadot X =0 ~~ .}}
    This multiplet exists in all Lagrangian field theories, which
    have no FI-terms and for which the K\"ahler form of their target space is exact.
\item{2.} New minimal supergravity \refs{\AkulovCK,\SohniusTP} is based on
the $R$-multiplet (see e.g.\ section 7 of \GatesNR)
    \eqn\Rmu{\eqalign{
    &\bar D^\alphadot \CR_{\alpha\alphadot} = \chi_\alpha  \cr
    &\bar D_\alphadot \chi_\alpha = \bar D_\alphadot \bar \chi^\alphadot - D^\alpha \chi_\alpha =0 ~~ .}}
    This multiplet exists in all field theories that have a global $U(1)_R$ symmetry,
    even in the presence of FI-terms or a target
space with a nontrivial K\"ahler form.  The conserved $U(1)_R$
current $j^{(R)}$ is the bottom component of
$\CR_{\alpha\alphadot}$.
\item{3.} 16/16 supergravity \refs{\GirardiVQ\LangXK-\SiegelSV} is based on a
larger multiplet \refs{\MagroAJ,\KomargodskiRB}, called the
S-multiplet,
    \eqn\Smu{\eqalign{
    &\bar D^\alphadot \CS_{\alpha\alphadot} = D_\alpha X +\chi_\alpha\cr
    &\bar D_\alphadot X =0\cr
     &\bar D_\alphadot \chi_\alpha = \bar D_\alphadot \bar \chi^\alphadot - D^\alpha \chi_\alpha =0~~ .}}
    This multiplet exists in all known Lagrangian field theories.
It is particularly important in theories without a global $U(1)_R$
symmetry, which have FI-terms or a target space with a nontrivial
K\"ahler form.

\noindent Additional multiplets and the relations between them were
discussed in \refs{\MagroAJ\KuzenkoAM-\KuzenkoNI}.

Each of these
multiplets includes the energy momentum tensor $T_{\mu\nu}$ ($6$
independent bosonic operators) and the supersymmetry current
$S_{\mu\alpha}$ ($12$ operators) as well as additional operators.
Some of these operators can be interpreted as brane currents for
branes of various dimensions.\foot{It is common to refer to them as
``brane charges'' or ``central charges.''  These terms are
misleading. First, when these ``charges'' correspond to extended
branes (not to 0-branes), the corresponding charge is
infinite and only the charge per unit brane volume is meaningful.
Therefore, we prefer to discuss the corresponding currents and the
charges \chargedef, which are finite.  Second, these charges are not
central -- they do not commute with the Lorentz group.}

In section 4 we argued that all such brane currents must be coupled
to gauge fields. In fact, for brane currents that are in the
supersymmetry multiplet this conclusion follows from imposing
supersymmetry -- since $T_{\mu\nu}$ and $S_{\mu\alpha}$ are coupled
to gauge fields, their superpartners must be as well.

There are a few important remarks that we should make about the
brane currents that appear in the super-current multiplets:
\item{1.}  The FZ-multiplet \FZmu\ includes the conserved, complex 2-brane
current $dx$ where $x$ is the lowest component of $X$ in \FZmu\ (recall that our conventions for currents \chargedef\ are dual to the more standard ones).  For
example, in Wess-Zumino models $x= 4 W -{1\over 3} \bar D^2 K $,
where $W$ is the superpotential and $K$ is the K\"ahler potential.
The tension of domain walls (2-branes) is constrained by this
current \DvaliXE.  In supergravity this current must be gauged.  The
corresponding gauge field is a 3-form $A^{(3)}$.  If $x$ is well
defined, the supergravity Lagrangian depends only on its field
strength $F^{(4)}=dA^{(3)}$, which couples to $x$. Integrating out
$F^{(4)}$ eventually leads to the famous $-|W|^2$ in the
potential.  This interpretation of the $|W|^2$ term as arising from
integrating out the gauge field of 2-brane currents is similar to
the relation between the Romans mass in ten dimensional IIA
supergravity and the gauge field of 8-branes.
\item{2.}  The $R$-multiplet \Rmu\ includes a conserved $2$-form
string current $Z$ in the $\theta$ component of $\chi_\alpha$.  For
example, in Wess-Zumino models $\chi_\alpha = \bar D^2 D_\alpha K$
where $K$ is the K\"ahler potential and the string current
$Z=\omega$ is the pull back of the K\"ahler form to spacetime.  In
theories with an FI-term $\chi_\alpha= -4\xi W_\alpha$ where $\xi$ is
the FI-term and $W_\alpha$ is the superfield including the field
strength $F$.  The corresponding string current is $Z= \xi F$.  Note
that if this gauge field is $U(1)$ rather than $\R$ and it is
coupled to dynamical magnetic monopoles, then its field strength $F$
is not closed and therefore the defining equation \Rmu\ is not
satisfied.\foot{If such a theory indeed exists, one might need to
use the more complicated multiplets discussed in
\refs{\MagroAJ\KuzenkoAM-\KuzenkoNI}.}
\item{3.} The $S$-multiplet \Smu\ includes both a
string current and a 2-brane current.

The interpretation of the brane currents in the various
supersymmetry multiplets leads to a new perspective on some of the
issues raised in \refs{\KomargodskiPC,\KomargodskiRB}.  The problem
with theories with nonzero FI-terms or with a non-exact K\"ahler
form is that their supersymmetry current algebra includes string
currents that cannot be improved to zero (see the discussion around
\chargedef). Therefore, the FZ-multiplet does not exist.

If the rigid theory has a global $U(1)_R$ symmetry, it has an
$R$-multiplet, which includes such a string current $Z$ and
therefore such theories can be coupled to supergravity using the
new-minimal formalism.  Even though both the $U(1)_R$ global current
$j^{(R)}$ and the string current $Z$ of the rigid theory are gauged,
the resulting supergravity theory has both a global $U(1)_R$
symmetry and a conserved global string current.  More explicitly,
the supergravity multiplet includes a one-form gauge field $A^{(R)}$
for the global $U(1)_R$ symmetry of the rigid theory, and a $2$-form
gauge field $B$  for the string current of the rigid theory.  These
two gauge fields couple through \eqn\BFnewmin{B\wedge d A^{(R)}~~.}
Therefore, the gauged $U(1)_R$ current is a linear combination of
the original the $U(1)_R$ current of the rigid theory $j^{(R)}$ and
$dB$.  Similarly, the gauged string current is a linear combination
of the original string current $Z$ and the field strength $F^{(R)}=d
A^{(R)}$.  Therefore, the supergravity theory has global conserved
currents which can be taken to be $j^{(R)}$ and $Z$ (or $F^{(R)}$ and
$H=dB$).  Note that the gauge fields $A^{(R)}$ and $B$ do not
correspond to massless propagating degrees of freedom and can be
integrated out.  The resulting on-shell theory still has global
conserved currents.  If we accept the arguments in section 4, we
conclude that such rigid theories cannot arise in consistent models
of quantum gravity \KomargodskiPC\ (see also \refs{\KuzenkoYM,\DienesTD}).

Even though the FZ-multiplet does not exist in these cases one can
still use the old-minimal formalism.  In fact, it is well known
that these two formalisms are dual to each other \FerraraDH\ (see however \LambertDX) and
therefore we can use either one.  As is common in the old minimal
formalism, one uses compensator fields.  In the problematic
situations the FZ-multiplet is either not gauge invariant or not
globally well defined.  This is taken care of by using compensator
fields, which are also non-gauge invariant or fail to be globally
well defined. This is possible only when the theory has a global
$U(1)_R$ symmetry and ultimately rests on the fact that the
$R$-multiplet exists.

This understanding naturally leads us to examine rigid theories
without a global $U(1)_R$ symmetry but with nontrivial string
currents.  Here we must use the $S$-multiplet and the corresponding
16/16 supergravity.  As pointed out in \SiegelSV\ for pure 16/16
supergravity and in \KomargodskiRB\ for the case with matter, this
theory has an alternate interpretation.  It can be interpreted as
standard minimal supergravity theory in which the rigid matter
system is enlarged by adding to it a chiral superfield $\tau$.  The
real part of the scalar in $\tau$ is dual to the $2$-form gauge
field $B$, which couples to the string current.  Note that unlike
the situation in the new-minimal supergravity, which we discussed
above, here $B$ or its dual scalar correspond to a massless
propagating particle.  With this interpretation of the theory the
additional chiral superfield $\tau$ removes the problem with the
coupling to supergravity \KomargodskiRB.  A simple way to understand
it is to note that with $\tau$ the string current can be improved to
zero; i.e.\ the $2$-form current is exact.  In theories with an
FI-term, that term becomes field dependent \DineXK; more precisely, the FI-term can be absorbed by shifting $\tau$ such that the theory does not really have an FI-term.  Similarly, in
theories with nontrivial cycles, the cycles become contractible in
the higher dimensional moduli space that includes $\tau$
\KomargodskiRB.

So far we viewed the additional field $\tau$ as infinitesimal.
However its global properties can be important.  Consider a $U(1)$ gauge theory with an FI-term which is eliminated through the coupling to $\tau$.  The complex field $\exp(2\pi i \tau) $ transforms linearly under the $U(1)$ gauge symmetry, but we have not yet discussed its charge.  It can be an arbitrary integer $p$.  In that case, the nonzero value of $\exp(2\pi i \tau) $ Higgses the $U(1)$ gauge symmetry down to $\Z_p$ and the system can have $\Z_p$ strings.  For example, this situation with field dependent FI-terms arises in string constructions with anomalous $U(1)$ \DineXK.  The anomaly is canceled through a Green-Schwarz term $B\wedge F$, where $F$ is the $U(1)$ field strength and $B$ is dual to ${\rm Re} \tau$.  The coefficient of this Green-Schwarz term is proportional to the sum of the $U(1)$ charges in the problem and thus determines the value of the integer $p$.

An interesting situation arises when the rigid
theory has several $U(1)$ gauge fields $V^i$ with transformation
parameters $\Lambda^i$ and FI-terms $\xi_i$.  The chiral superfield
$\tau$ is added to the rigid theory so that under gauge
transformations it transforms as \eqn\tautra{\tau \to \tau + i
\Lambda_i \xi^i~~.} When some of the gauge groups are compact, i.e.\
$\Lambda_i \sim \Lambda_i + 2\pi i$, the global structure of $\tau$
becomes important.  In this case if $\xi_i$ are generic, the gauge
transformation rule \tautra\ is incompatible with any periodicity of
$\tau$.  Similarly, we can consider a rigid theory with a
complicated target space with several two-cycles.  The addition of
$\tau$ through $K \to K + i(\tau-\bar \tau)$ \KomargodskiRB\ might
be incompatible with all the necessary K\"ahler transformations
associated with all the two-cycles\foot{We thank N.~Nekrasov for a
useful discussion about this point.}.

One way to address this question is to constrain the maps from
space-time to the target space.  In the gauge theory case we could
simply declare that all the gauge groups are $\R$ rather than
$U(1)$.  And in the $\sigma$-model case we can follow section 2 and
\SeibergQD\ and restrict the maps such that the wrapping of various
cycles is compatible with the existence of $\tau$.  This amounts to
adding an additional discrete gauge symmetry to the system. This
modification of the rigid theory by adding $\tau$ and an appropriate
modification of its global structure makes the coupling to
supergravity possible. However, the discussion in section 4 points
us in another direction.

The issue with the global behavior of $\tau$ arises when the
addition of $\tau$ to the rigid theory does not eliminate all its
string currents.  The addition of the discrete gauge symmetry in the
previous paragraph, which restricts the maps from spacetime to the
target space, induces a global $U(1)$ symmetry which shifts $\tau$
by an arbitrary real constant.  Although this is perfectly
consistent with perturbative supergravity, it violates the
conjectures of section 4.

Therefore, the only way to weakly couple such systems to
supergravity, while respecting the three conjectures of section 4, is
the following.  For generic $\xi_i$ and for generic cycles, we need
to add a separate $\tau_i$ field for every gauge group and for every
two-cycle.  In other words, all FI-terms must be field dependent and
all the K\"ahler moduli must be massless dynamical fields.\foot{This
observation extends the condition about massless moduli in
\KomargodskiRB\ to many additional moduli.  It has the consequence
that in a supersymmetric string construction with a continuously variable
ratio between low energy scales and the Planck scale, many of the moduli cannot be stabilized in a manner consistent with SUSY.}

\subsec{Intrinsically gravitational theories}

In the previous subsections we studied a quantum field theory weakly
coupled to supergravity, with a continuously variable coupling
parameter. Assuming the conjectures of section 4., we concluded that
all FI-terms are field dependent (which means they can be removed),
and no two-cycles are present in the target space.

Here we consider situations in which the separation between the
rigid field theory and gravity is not possible.  In particular, we
are interested in theories with FI-terms or with target spaces that
have nontrivial cycles of the order of the Planck scale.

A careful analysis of the component Lagrangian shows that FI-terms
must be quantized\foot{We use notation ${1\over G_N} =M_{Planck}^2 =
{8\pi \over \kappa^2 }= 8\pi M_P^2 $; i.e.\ $M_P$ is the reduced
Planck mass.} \SeibergQD\ (see also \DistlerZG)
 \eqn\xicont{\xi = 2 N M_{P}^2    \qquad , \qquad N \in \Z ~. }
Similarly, if the target space includes nontrivial two-cycles, their
periods must be rational.  For example, if the target space includes
a $\CP^1$ with the metric \eqn\CPOmet{ds^2 = f_\pi^2 {d \Phi d \bar
\Phi \over (1 + |\Phi|^2)^2} ~~,} $f_\pi$ must satisfy
 \eqn\fpicon{f_\pi^2 = {2N \over  p} M_{P}^2  \qquad , \qquad p, N \in \Z }
and the integer $p$ is related to a discrete $\Z_p$ gauged R-symmetry.
(The case $p=1$ was studied in \WittenHU.)

The quantization conditions \xicont\fpicon\ were derived in
\SeibergQD\ by focusing on the gravitino and its transformation
laws. These are gauge transformations in the case with FI-terms, and
K\"ahler and $\Z_p$ transformations in the case with two-cycles.
Here we will re-derive and interpret these results from the
perspective of our discussion -- using the underlying string
currents and the coupling to a $2$-form gauge field.

The new-minimal formalism is particularly useful if one wants to
track the remaining discrete gauge symmetries.  Let us start with a
model with an $R$-symmetry such that it is straightforward to use
this formalism.  As explained around \BFnewmin, even though the
R-symmetry is gauged, the theory still has a global R-symmetry whose
current can be taken to be $j^{(R)}$. Next we try to add to the
Lagrangian terms that explicitly break this symmetry.  However,
since a linear combination of $j^{(R)}$ and $H=dB$ is gauged, we
must preserve that linear combination.  In other words, the
conservation of $H$ must also be violated.  This can be done only if
the one-form gauge symmetry of $B$ is compact.  Then, as explained
in section 2, there exist local operators $\CO_N$ around which
$\int_{\S^3} H =2\pi N \not=0$ ($N\in \Z$).  These operators carry
charge $N$ under the global symmetry. Using $\CO_N$ and R-breaking
operators constructed out of matter fields, we can explicitly
violate the conservation of both $H=dB$ and $j^{(R)}$, while
preserving their gauged linear combination.

Similarly the theory has two conserved string currents, $Z$ and
$F^{(R)}=dA^{(R)}$.  One linear combination of them is gauged.  We
can preserve that linear combination, while violating the
conservation of $Z$ and $F^{(R)}$, by adding to our system particles
magnetically charged under $A^{(R)}$, and having suitable properties
under $Z$.

The upshot of all this is that if the gauge symmetries of $A^{(R)}$
and of $B$ are compact, and the appropriate magnetic objects are
present, both the global R-symmetry and the conservation of the
global string current can be violated.  Of course, since these
symmetries are compact, their corresponding charges are quantized.
This leads to the quantization conditions \xicont\fpicon.

The same conclusion can be reached using the old-minimal formalism.
Here the point is that the compensators are not gauge invariant and
not K\"ahler invariant.  Normally, the compensator fields appear only in front of the exponential of the K\"ahler potential and in front of the superpotential.  However, if the quantization conditions
\xicont\fpicon\ are satisfied, we can add to the Lagrangian terms with other dependence on the compensators such that they explicitly
break the global R-symmetry.

In conclusion, the problem with FI-terms and with nontrivial
two-cycles can be traced back to the existence of strings in the
rigid model that we are trying to couple to SUGRA. With FI-terms the
quantization condition \xicont\ guarantees that magnetic monopoles
can be added to the system such that the strings are unstable.  With
nontrivial two-cycles the condition \fpicon\ guarantees that one can
couple the strings to a $\Z_p$ gauge theory and violate the string
current conservation.

\bigskip
\centerline{\bf Acknowledgements} We thank N.~Arkani-Hamed, O.~Aharony, M.~Dine, T.~Dumitrescu, J.~Distler, M.~Douglas, D.~Freed, Z.~Komargodski, J.~Maldacena,
G.~Moore, N.~Nekrasov, J.~Polchinski, M.~Rocek, S.~Shenker,
A.~Strominger, and E.~Witten for useful discussions. The work of TB was supported in part by DOE grant
DE-FG03-92ER40689 and in part by the Ambrose Monell Foundation at
the Institute for Advanced Study. The work of NS is partially
supported by DOE grant DE-FG02-90ER40542. Any opinions, findings,
and conclusions or recommendations expressed in this material are
those of the author(s) and do not necessarily reflect the views of
the funding agencies.

\appendix{A} {Generalizations of the Lagrangian description of the $\Z_p$ gauge theories}

It is easy to extend the $BF$-theory to $\CN=1$ supersymmetry.  The gauge field $A$ belongs to a vector superfield $\CV$ with the gauge symmetry
\eqn\supergau{\CV \to \CV + \Lambda + \Lambda^\dagger  ~~.}
The $2$-form $B$ belongs to a chiral spinor superfield $\CB_\alpha$ (satisfying $\bar D_\alphadot \CB_\alpha= 0$) with the gauge symmetry
\eqn\Psigauge{\CB_\alpha \to \CB_\alpha + \bar D^2 D_\alpha L}
for arbitrary real $L$.  The gauge invariant field strength $H=dB$ is embedded in the real linear superfield\foot{A real linear superfield $\CH$ is characterized by $\CH=\CH^\dagger$ and $ D^2 \CH=0$ (and using the reality also $\bar D^2 \CH=0$).} $\CH=  D^ \alpha \CB_\alpha + h.c.$ which is invariant under \Psigauge.

Then the $BF$-Lagrangian \BFlaga\ is included in
\eqn\BFSus{{p\over 2\pi} \int d^4 \theta \CH \CV= {p\over 2\pi}  \int d^2 \theta \CB_\alpha W^\alpha + h.c.}
which is invariant under the two gauge symmetries \supergau\Psigauge.

Our discussion of the $BF$-theory has an obvious generalization to an arbitrary $q$-form gauge field $A$ in $d$ dimensions \MaldacenaSS. The $\Z_p$ gauge theory is described at low energies by two $U(1)$ gauge fields: a $q$-form $A$ and a $d-q-1$-form $B$ with the Lagrangian
\eqn\BFlowa{{i p\over 2\pi} B \wedge dA~~.}
Note that this theory is invariant under $q \to d-q-1 $
which exchanges $A$ and $B$.

A particularly interesting special case is $q=1$ in three
dimensions, where both $A$ and $B$ are one-forms.  In this case we
find a $U(1)\times U(1)$ Lagrangian description of the three
dimensional $\Z_p$, level $k$, Chern-Simons gauge theory of
\refs{\DijkgraafPZ,\FreedBN} (see also the recent discussions
\refs{\BelovZE,\KapustinHK}) \eqn\zpcst{{i p\over 2\pi} B \wedge dA
+ {i k \over 4\pi} A \wedge dA~~.} As in \Elecop, the theory has
$p^2$ line operators \eqn\threedli{\exp\left( in_A \oint_{\Sigma_1}
A + in_B \oint_{\Sigma_1} B\right) } labeled by $n_A, n_B=0,1,...p$.

\appendix{B}{Topology changing operators in the chiral Lagrangian}

In this brief appendix, we note that our method of describing
operators that violate topological symmetries in terms of boundary
conditions, can easily be extended to other $\sigma$-models.

Consider for example the $SU(n)$ chiral model.  The dynamical field is a
group element $\Sigma \in SU(n)$ and the model has an  $SU(n ) \times SU(n ) $ symmetry.  We compactify space to $\S^3$ and then the states of the system are labeled by the integer quantum number $\pi_3(SU(n)) = \Z$ and the corresponding current is
\eqn\Bcurrent{B  \sim {\rm tr}\ (\Sigma^{\dagger} d \Sigma )^3 ~~,}
where the cubic product is a wedge product.  As is well known, in the context
of QCD, this current should be interpreted as the baryon current, and the
corresponding solitons are baryons \refs{\SkyrmeVQ\FinkelsteinHY-\WittenTX}.

We can define a local operator $\CO_k$ that creates $k$ baryons.  We
remove a point from spacetime and impose that $\Sigma$ on its
surrounding $\S^3$ is associated with $k$ wrappings in $\pi_3(SU(n))
= \Z$.  $\CO_{k=1}$ is an interpolating field for a single baryon.
We can add $\CO_k$ to the Lagrangian to describe the effects of UV
physics that violates baryon number but preserves a $\Z_k$ subgroup
of it.

Another example of the general phenomena we have explored is the
observation that the chiral Lagrangian for an $O(n)$ gauge theory
coupled to $n$ chiral fermions in the vector representation, has
$\Z_2$ strings arising from $\pi_2 (SU(n)/O(n) ) = \Z_2$ \WittenTX.
We can construct a string creation operator by excising a one
dimensional line from space-time and imposing boundary conditions
around it. Locally the line is surrounded by $\R\times \S^2$ and we
impose that on the $\S^2$ the fields are associated with the
non-trivial homotopy class.

\appendix{C}{Lattice gauge theories}

There are higher form $\Z_p$ invariant gauge theories on any $d$ dimensional cubic lattice.  In fact, our discussion generalizes rather simply to any finite Abelian group and any $d$ dimensional simplicial complex. A $q$-form
$\Z_p$ gauge theory is an assignment of an integer $L\mod p$ to
each $q+1$ dimensional hypercube on the lattice. (Recall our convention labeling higher form gauge symmetry by the dimension of the gauge parameter.) For $q = -1,0$ the
hypercube is a point or a link. To simplify the notation, in this
appendix we will restrict attention to $q=-1$ and $d=4$, which is
generally called the case with {\it global} $\Z_p$ symmetry. Our
considerations are easily generalized to arbitrary $q$ and $d$. In
particular, our conclusion that, in the symmetric phase, the global
$\Z_p$ symmetry is realized as the global limit of an IR gauge
symmetry is true for general $d$.

The lattice partition function is the sum over all $L (x) $
of\foot{We limit ourselves to nearest neighbor interactions. More
complicated local actions which couple more points are possible.}
\eqn\Cone{e^{\sum_{x,\mu } S[d_{\mu}L(x)]} ~~,} where the sum in the
exponent is over all elementary links. $d_{\mu}$ is the lattice
finite difference operator in the direction labeled by $\mu$. Correlation functions are expectation values of local operators of the form
\eqn\Ctwo{e^{ {{2\pi i}\over {p}} n_k L (x)}~~,~~ n_k=0,...,p-1  ~~.}
(For higher values of $q$ the observables are gauge invariant products of elements like \Ctwo.)
$\Z_p$ invariance would imply that only
correlation functions with  $\sum n_k = 0\mod p$ are
non-vanishing, but there are values of the parameters where the
$\Z_p$ symmetry is spontaneously broken\foot{If we were dealing with
$q \ge 0$, this language would be slightly misleading, but
conventional. The local symmetry is never spontaneously broken.  This phase is called the Higgs phase for $q \ge 0$.}.

Next, we apply ordinary electric magnetic duality to this system and replace the variables $L(x)$ on the sites with plaquette variables.  We define  $e^{\tilde{S} [z]}$ to be the $\Z_p$ Fourier transform of
$e^{S[z]}$. We use the same letter for the arguments of the two
functions because the group $\Z_p$ is self-dual. Using lattice
integration by parts one easily shows that the partition function is
given by \eqn\Cthree{\sum_{K_{\mu} (x)} e^{\sum_{x, \mu} \tilde{S}
[K_{\mu}(x)]}  \prod_x \delta [ d_{\mu} K_{\mu} (x) ] ~~. } The
$\Z_p$ valued vector $K_{\mu}$ must be conserved. The $\delta$
function in this equation is a $\Z_p$ valued Kronecker $\delta$ .

For correlation functions, conservation of $K_{\mu}$ fails on those
points where there are local operators. We can solve the
conservation constraint by writing \eqn\Cfour{K_{\mu} =
\epsilon^{\mu\nu\rho\sigma} d_{\nu} L_{\rho\sigma} + \Delta ~~.}
The $\Z_p$ valued $2$-form $L_{\mu\nu}$ really lives on the dual lattice, but we
use the Levi-Civita symbol to assign it to a lattice plaquette.
$\Delta$ is a sum of Dirac string contributions, emanating from each
point on the original lattice where conservation failed. The
correlation functions are invariant under $L_{\mu\nu}
\rightarrow L_{\mu\nu} + d_{\mu} L_{\nu} - d_{\nu}
L_{\mu}$.  The Dirac strings must either go to infinity, or
connect points with opposite charge. Any choice of
strings satisfying these conditions is equivalent to any other
choice, because the difference between them defines a $K_{\mu}$
which is conserved, and can be written in terms of
$L_{\mu\nu}$.\foot{On simplicial complexes with more complicated
topology we would have to include variables describing closed but
inexact forms.}  We see that a theory with global $\Z_p$ symmetry is
dual to a theory with a one-form $\Z_p$ gauge symmetry. More
generally, a theory with a $k$-form gauge symmetry is dual to one
with $(d - 4 - k)$-form gauge symmetry. Correlation functions of the
original fields involve expressions with non-local Dirac strings,
when they are written in terms of the dual fields.

The action $S$ depends on $p$ parameters, and in some extreme
parameter ranges it is large. We can then approximate the partition
sum by finding the configuration that minimizes $S$ and considering
a dilute gas of small localized fluctuations around it. This is
actually the first term in a convergent expansion and defines a weak
coupling phase of the lattice theory.  Similarly, there is a
different range of parameters in which $\tilde{S}$ is large, and we
have a strong coupling phase.  Both phases have a mass gap finite in
lattice units.  The IR physics is completely topological and is
described by a $BF$-Lagrangian. For the weak coupling, spontaneously
broken phase, the IR Lagrangian has a 0-form potential and a
$3$-form potential. The latter couples to the domain wall
excitations of the spontaneously broken phase. In the strong
coupling, symmetric, phase the IR Lagrangian has a $2$-form
potential $B$ and a $1$-form potential.  The Dirac
strings that attach to the charged local operators, are realized as
Wilson lines of the $1$-form potential.  They can be shifted at
will, as long as there are no sources coupling
to the $2$-form.

We conclude that in the symmetric phase, the
global symmetry of the original model is realized as the global
transformation that is the limit of a $\Z_p$ $0$-form (Maxwell) gauge
symmetry, coupled by the topological Lagrangian to a $1$-form
(Kalb-Ramond) gauge potential.  If we take the continuum limit, the
tension of the co-dimension $2$ objects that couple to
$B_{\mu\nu}$ goes to infinity, so we can always ignore the fact
that the global symmetry is the global part of a gauge group.

We want to emphasize that we are {\it not} saying that every $\Z_p$
one form gauge symmetry has a global symmetry associated with it. In
QFT, exact global symmetries are not emergent -- at best there can be accidental approximate global symmetries.  However gauge symmetries
are often emergent. We have
shown that a particular class of UV regulators, the lattice,
associates a one form $\Z_p$ gauge symmetry to every system with a
global $\Z_p$ symmetry. In the IR limit of the symmetric phase, there
is an emergent $0$-form gauge symmetry, and the original global
symmetry becomes the global limit of these gauge transformations in
the IR.  We can run this argument backward only if we have the full
UV complete $1$-form lattice gauge theory.

It is illuminating to restate this using our continuum formalism.   We start with a continuum $\Z_p$ gauge theory, ${\cal
L} = {ip \over 2\pi} B \wedge dA$ and would like to introduce local operators. We do that by adding to the system charged fields $\chi(x)$ and study the gauge invariant operators
\eqn\chiBF{\chi (x) e^{ i \int_x^{\infty}A} ~~.}
These are not invariant under gauge transformations that approach
a constant at infinity.  Therefore, the operators \chiBF\ can be interpreted as carrying charge under a global symmetry.

Conversely, starting with a linearly realized global
$\Z_p$, acting on a field $\psi (x) $, we can write $\psi (x) =
\sqrt{\psi^* \psi} e^{i \int_x^{\infty}\ V}, $ where $V$ is a one
form and the integral is taken along some contour.  Adding to the action $e^{ {ip\over 2\pi} \int\ B \wedge dV}$ we see that $B$ acts as a Lagrange multiplier setting $dV=0$ and therefore the expression for $\psi(x)$ is independent of local changes in the contour. We have thus introduced a fake gauge
invariance, and in this formalism the global symmetry is realized as
the global part of the gauge group --
the group of gauge transformations that act as constants at
infinity, modulo those which act as the identify at infinity.

These observations may be relevant to our speculations about
discrete gauge symmetries in section 4.2.  If gravitational effects
force the co-dimension 2 brane tension to be finite, we would be
able to prove that all discrete symmetries are gauged.

For many forms of the action $S$, appropriate values of $d$, and
large enough $p$, there is another phase of the system, which
illustrates some of the principles we have expounded in the text.
This is well known in the literature \refs{\ElitzurUV,\UkawaYV}, so
we will only sketch the results. Start from the dual form of the
$\Z_p$ theory and use the coset construction of $\Z_p$ in terms of
$\Z$ to write this as a $\Z$ valued gauge theory with matter fields
of charge $p$. Then use the Poisson summation formula to write the
sum over $\Z$ valued fields in terms of an integral and another sum.
For $q=0$ and $d=4$, the result is a lattice version of
QED, coupled to electric and magnetic charges.  The electric charges
are $p$ times the fundamental unit of charge, while the monopoles
have the minimal Dirac quantum.  There is often an intermediate
Coulomb phase, where the IR physics is that of a free photon, and
the electric and magnetic particles have mass of order the inverse
lattice spacing. The weak and strong coupling phases can be viewed
as transitions to states where either electric or magnetic charges
are condensed. The charges supply the relevant ends of strings in
the weak and strong coupling phases.  The Coulomb phase has no
stable strings and is an illustration of the instability of
quasi-Aharonov-Bohm strings \PolchinskiBG.

\listrefs

\end